\documentclass{article}

\usepackage{PRIMEarxiv}

\usepackage[utf8]{inputenc} 
\usepackage[T1]{fontenc}    
\usepackage{hyperref}       
\usepackage{url}            
\usepackage{booktabs}       
\usepackage{amsfonts}       
\usepackage{nicefrac}       
\usepackage{microtype}      
\usepackage{lipsum}
\usepackage{fancyhdr}
\usepackage{amsmath}
\usepackage{graphicx}       
\usepackage{subcaption} 
\usepackage{natbib}
\providecommand{\doi}[1]{\href{https://doi.org/#1}{doi: #1}}
\graphicspath{{media/}}     

\title{Estimation of Elastic Parameters with Guidance-based Diffusion model
}

\author{
  Anjali Dixit, Francesco Brandolin, Tariq Alkhalifah \\
  King Abdullah University of Science and Technology (KAUST) \\
  Thuwal, Kingdom of Saudi Arabia\\
  \texttt{\{anjali.dixit, francesco.brandolin, tariq.alkhalifah\}@kaust.edu.sa} \\
}

\begin{document}
\maketitle


\begin{abstract}
Elastic parameters are fundamental rock properties for reservoir characterization, but their reliable estimation from angle-stack seismic data remains challenging due to strong nonlinearity and imperfect physical modeling. Conventional deterministic approaches based on linearized Zoeppritz approximations yield a single point estimate and cannot quantify solution uncertainty, while probabilistic methods are computationally expensive. To address these limitations, we present a workflow for elastic parameter inversion from angle-stack seismic data using a guided diffusion model as an implicit prior over the joint distribution of P-wave velocity, S-wave velocity, and density. The diffusion model is trained in an unsupervised manner on benchmark datasets and well-log-derived synthetic models, so that it learns the non-Gaussian statistical coupling among the three elastic parameters. For guidance, we employ Diffusion Posterior Sampling (DPS), which approximates the likelihood function through a forward operator based on the Aki-Richards approximation and injects data-consistency gradient corrections at each reverse diffusion step, allowing sampling from a posterior conditioned on the misfit between observed and modeled angle-stack data. We evaluate the framework on two datasets: the 2D Otway synthetic elastic model and field data from the Poseidon field, NW Shelf, Browse Basin, Australia, comparing it against two conventional baselines: LSQR least-squares inversion and ADMM-based inversion with total variation regularization. Quantitative comparisons confirm that the diffusion-based framework recovers sharper lithological contrasts and geologically more realistic elastic profiles. Uncertainty quantification is achieved by generating multiple independent posterior realizations through repeated reverse diffusion runs, producing spatially resolved uncertainty maps for each elastic parameter.
\end{abstract}


\section{Introduction}

Estimating elastic parameters, namely P-wave velocity ($V_P$), S-wave velocity ($V_S$), and density ($\rho$) is of fundamental importance in seismic reservoir characterization, as these parameters carry direct information about lithological variations, fluid content, and pore pressure \citep{castagna1993offset, christensen1996poisson, grana2016bayesian}. Accurate estimation of elastic properties enables the discrimination between gas-saturated and brine-saturated sands \citep{castagna1993offset, smith2003geophy}, the identification of lithofacies boundaries \citep{avseth2010quantitative, grana2017seismic}, and the monitoring of subsurface fluid substitution during production, which have critical role in hydrocarbon exploration and carbon capture and storage (CCS) applications \citep{lumley2001time, arts2004monitoring, zhang2022monitoring}. Prestack amplitude-variation-with-angle/offset (AVA/AVO) inversion is the standard practice for estimating elastic parameters from seismic reflection data. In this approach, angle-dependent reflection amplitudes recorded at the surface are inverted to recover the contrasts in $V_P$, $V_S$, and $\rho$ at subsurface interfaces. Conventional AVA inversion relies on linearized approximations of the Zoeppritz equations \citep{aki2002quantitative}, most notably the three-term Aki-Richards approximation and the subsequent simplification by \citet{shuey1985}, to establish an analytically tractable linear relationship between the angle-dependent reflection coefficients and the elastic parameter contrasts. These linearized forward operators form the basis of deterministic inversion methods, which seek a single best-fit model by minimizing a weighted least-squares misfit between observed and modeled seismic amplitudes, often supplemented with regularization to stabilize the ill-posed problem \citep{tikhonov1977}. While computationally efficient, deterministic methods yield a single point estimate of the elastic model and provide no quantification of the solution non-uniqueness inherent to the seismic inverse problem \citep{tarantola2005}.

Another approach to solve this inverse problem is to cast it within a Bayesian probabilistic framework \citep{tarantola2005}. By combining the prior distribution with the likelihood function, the posterior distribution of the subsurface variables can be approximated as:
\begin{equation}
p(\mathbf{m} \mid \mathbf{d}) \propto p(\mathbf{d} \mid \mathbf{m}) p(\mathbf{m}),
\end{equation}
where $p(\mathbf{d} \mid \mathbf{m})$ is the likelihood function representing the fidelity of the predicted seismic data to the observations under an assumed noise model, and $p(\mathbf{m})$ represents the prior distribution encoding geological knowledge about the subsurface. The posterior $p(\mathbf{m} \mid \mathbf{d})$ encapsulates all information about the model parameters consistent with both the observed data and the prior, thereby providing not only a best-fit estimate but also a complete characterization of uncertainty arising from the ill-posedness of the inverse problem. However, estimating the posterior distribution remains challenging for AVA inversion \citep{grana2022probabilistic}, primarily because the problem is high-dimensional and the prior distribution of realistic subsurface models is complex and non-Gaussian.

Under the simplifying assumptions of a linearized forward model and Gaussian prior distributions, \citet{buland2003bayesian} derived a closed-form analytical expression for the posterior, providing an efficient but restrictive solution valid only in the weak-contrast regime. To relax the Gaussianity assumption, \citet{grana2010probabilistic} extended the analytical framework to Gaussian mixture models, enabling the representation of multimodal posteriors; however, the analytical posterior becomes intractable for mixture priors and requires stochastic sampling to be evaluated. Consequently, Markov Chain Monte Carlo (MCMC) methods have been widely adopted as standard samplers for the posterior in seismic inversion (\citealp{mosegaard1995}), as they make no parametric assumption about the posterior shape and can handle nonlinear forward models and non-Gaussian priors (\citealp{aleardi2019}). Nevertheless, MCMC suffers from prohibitive computational cost in high-dimensional model spaces, poor convergence in the presence of multimodality, and limited scalability to 2D and 3D problems. Geostatistical approaches such as sequential Gaussian simulation and iterative geostatistical seismic inversion offered spatially correlated posterior realizations consistent with well-log statistics (\citealp{haas1994geostatistical, azevedo2018geostatistical, azevedo2019geostatistical}), yet at the expense of even higher computational overhead and without a tractable analytical expression of the posterior (\citealp{azevedo2018geostatistical}).

Recently, generative modeling approaches have emerged as effective tools for learning data-driven priors, with successful applications in geophysical inverse problems (\citealp{wang2023prior, chen2025unsupervised, ravasi2025}). Score-based diffusion models, in particular, have attracted significant attention as implicit priors owing to their ability to capture complex, non-Gaussian distributions of subsurface models learned directly from training data. \citet{chung2022improving} introduced Diffusion Posterior Sampling (DPS), a general framework that conditions a pretrained unconditional diffusion model on observed data by injecting likelihood-gradient corrections at each reverse diffusion step, without requiring task-specific retraining. Building on this paradigm, several studies have demonstrated the viability of diffusion-based priors across a range of seismic inverse problems. In the context of elastic parameter recovery, \citet{taufik2024} trained an unsupervised diffusion model to capture the joint distribution between elastic parameters (i.e., $V_P$, $V_S$, and $\rho$) and incorporated it as an adaptive plug-and-play regularizer within multi-parameter elastic inversion, demonstrating improved parameter ratio realism and convergence compared to conventional regularization schemes. \citet{taufik2025} further extended this framework by embedding a wavenumber continuation process into the diffusion-based regularization, enabling a coarse-to-fine recovery of elastic parameters consistent with the multi-scale nature of the inversion. For seismic impedance inversion, \citet{liu2024} proposed a conditional diffusion probabilistic model that reformulates the inversion as a Markov process with latent variables, using variational inference to reduce solution non-uniqueness and avoid the mode collapse issues common in adversarial training. More recently, \citet{zhao2025} proposed a latent diffusion framework for acoustic impedance inversion operating in a compressed latent space, significantly reducing computational cost while maintaining geological realism through a lightweight wavelet-based conditioning module \citep{zhao2025}. \citet{jin2026} proposed a DPS-based acoustic impedance inversion method with 3D lateral constraints, enforcing spatial consistency across traces during the reverse diffusion process. \citet{ravasi2025} applied measurement-guided diffusion processes to seismic inversion, identifying model reparameterization to $[-1, 1]$ and the choice of training dataset as critical factors for success. \citet{shi2025} proposed an unsupervised posterior sampling framework for multiple seismic inverse tasks using a diffusion model with a novel noise schedule, enabling flexible posterior sampling without task-specific retraining. In the context of ensemble-based methods, \citet{tle2025} integrated diffusion model priors within an ensemble smoother for seismic inversion, improving geological realism of posterior realizations while maintaining data consistency.

In this work, we integrate a diffusion-based prior within a Bayesian-inspired inversion framework, using Diffusion Posterior Sampling (DPS; \citealt{chung2022improving}) to guide the reconstruction of elastic properties, specifically P-wave velocity, S-wave velocity, and density from angle-stack seismic data. The prior distribution encoded by the diffusion model is learned from a dataset of high-resolution elastic models, capturing the joint distribution of the three parameters. We employ DPS to approximate the observation likelihood from the angle-stack seismic data and combine it with the diffusion-based prior during the reverse diffusion process, guiding the solution toward high-probability regions of the posterior. In this work, we consider a single realization, resulting in a posterior-guided reconstruction rather than a full characterization of the posterior distribution. Although a single realization can provide an acceptable sample from the posterior, the stochastic nature of the reverse diffusion process inherently supports ensemble-based uncertainty quantification by generating multiple independent posterior realizations, which we briefly examine to assess the spread of the reconstructed elastic models relative to the analytical Bayesian posterior. Specifically, we quantify uncertainty by computing the posterior mean and standard deviation  over an ensemble of independent DPS realizations, and benchmark the resulting uncertainty estimates against the exact posterior covariance derived from the closed-form analytical Bayesian inversion of \citet{buland2003bayesian}, which serves as a tractable reference under Gaussian prior and linearized forward model assumptions. This comparison allows us to critically evaluate the calibration of DPS-based uncertainty and identify its limitations as a practical uncertainty quantification tool for elastic parameter inversion. The major contributions of this work are as follows
\begin{itemize}
    \item We present a diffusion-based framework for estimating subsurface 
    elastic parameters i.e., P-wave velocity, S-wave velocity, and density, 
    wherein diffusion posterior sampling (DPS) enforces data consistency at 
    each reverse diffusion step to achieve physics-guided posterior 
    reconstruction.

    \item We introduce a linearized AVA forward operator based on the 
    Aki-Richards approximation as the likelihood guidance term within the 
    DPS framework, enabling physically consistent data-driven inversion 
    without task-specific retraining of the diffusion prior.

    \item We validate the proposed framework on both a 2D Otway 
    synthetic benchmark and field data from the Poseidon field, 
    NW-Shelf, Browse Basin, Australia, demonstrating superior 
    reconstruction accuracy over conventional least-squares (LSQR) 
    and ADMM-based total variation inversion baselines.

    \item We perform the uncertainty estimates for the DPS-based 
    framework against a closed-form analytical Bayesian inversion, 
    revealing the over-confident nature of DPS uncertainty bands 
    relative to the exact posterior covariance

\end{itemize}

\section{Theory}

\begin{figure}
\centering
\includegraphics[scale=0.11]{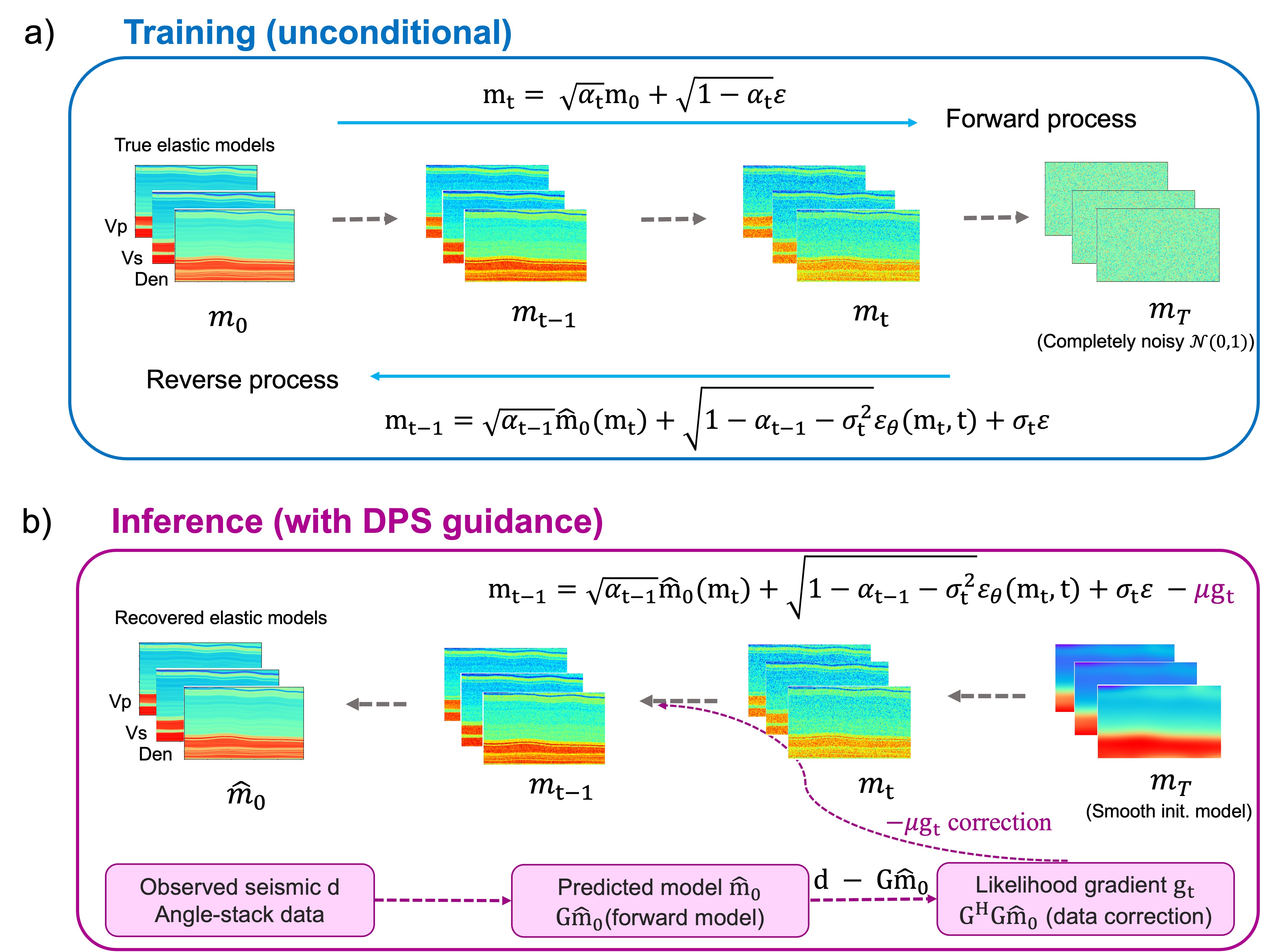} 
\caption{Schematic overview of the proposed workflow. a) Unconditional training stage: the forward diffusion process progressively corrupts true elastic models ($V_p$, $V_s$, density) into Gaussian noise, while the reverse process trains a denoising network to recover the elastic models from noise. b) Inference stage with DPS guidance: starting from a smooth initial model, the reverse diffusion process is guided at each step by a data-consistency correction ($-\mu \mathbf{g}_t$), computed from the residual between the observed angle-stack seismic data and the data predicted from the current model estimate via the forward operator ($\mathbf{G}\hat{\mathbf{m}}_0$).}

\label{figwf} 
\end{figure}

The proposed inversion framework consists of two stages: an unconditional training stage, in which a diffusion model learns a prior over the joint distribution of P-wave velocity, S-wave velocity, and density from a database of elastic models, and a guided inference stage, in which Diffusion Posterior Sampling (DPS) conditions the reverse diffusion process on the observed angle-stack seismic data. Figure~\ref{figwf} illustrates this workflow: the top panel shows the forward and reverse diffusion processes used during unconditional training, while the bottom panel shows how, at inference time, the predicted elastic model is passed through the AVA forward operator and compared against the observed data to compute a likelihood-gradient correction ($-\mu \mathbf{g}_t$) that is injected at each reverse diffusion step. The following subsections detail the mathematical formulation of each component of this workflow.

\subsection{Prestack AVA elastic parameters inversion}
Prestack seismic inversion seeks to recover the subsurface elastic
properties, i.e., P-wave velocity, S-wave velocity, and density from
angle-dependent reflection seismic data \citep{tarantola1984, buland2003bayesian}.
The key link between the elastic properties and the observed seismic data is the
angle-dependent PP reflection coefficient, which is approximated via
the linearized Aki-Richards expression \citep{aki2002quantitative}:
\begin{equation}
    R(\theta) \approx
    a(\theta)\, \frac{\Delta V_p}{\bar{V}_p}
    + b(\theta)\, \frac{\Delta V_s}{\bar{V}_s}
    + c(\theta)\, \frac{\Delta \rho}{\bar{\rho}},
    \label{eq:aki_richards}
\end{equation}
where $\Delta(\cdot)$ and $\bar{(\cdot)}$ denote the contrast and average
of a parameter across a reflecting interface, respectively, and the
angle-dependent weighting coefficients are\\

\begin{align}
    a(\theta) &= \frac{1}{2\cos^2\theta}, \\
    b(\theta) &= -4\frac{\bar{V}_s^2}{\bar{V}_p^2}\sin^2\theta, \\
    c(\theta) &= \frac{1}{2}\!\left(1 - 4\frac{\bar{V}_s^2}{\bar{V}_p^2}\sin^2\theta\right).
\end{align}
Following \citet{buland2003bayesian}, we adopt a log-domain
parameterization $\mathbf{m} = [\ln V_p,\, \ln V_s,\, \ln \rho]^T$,
so that the parameter contrasts reduce to finite log-differences,
i.e., $\Delta \ln V_p \approx \Delta V_p / \bar{V}_p$,
linearizing the reflectivity with respect to the model parameters.

The observed angle-stack data at angle $\theta_k$ are modeled as the
convolution of the reflectivity series $R(t, \theta_k)$ with a source
wavelet $w(t)$ \citep{goupillaud1961}:
\begin{equation}
    d(t, \theta_k) = w(t) * R(t, \theta_k) + n(t),
    \label{eq:forward_conv}
\end{equation}
where $n(t) \sim \mathcal{N}(0, \sigma_d^2)$ represents additive Gaussian
noise with variance $\sigma_d^2$.
Stacking contributions from $N_\theta$ angle stacks, the full forward
problem is expressed in compact matrix--vector form as
\begin{equation}
    \mathbf{d} = \mathbf{G}\mathbf{m} + \mathbf{n},
    \label{eq:forward_matrix}
\end{equation}
where $\mathbf{G} \in \mathbb{R}^{N_\theta n_z \times 3N}$ is the combined
wavelet-convolution and AVA forward operator,
$\mathbf{d} \in \mathbb{R}^{N_\theta n_z}$ is the vectorized angle-stack
data, and $\mathbf{n} \sim \mathcal{N}(\mathbf{0}, \sigma_d^2 \mathbf{I})$.
This operation is implemented using \texttt{pylops} \citep{ravasi2020pylops},
which provides matrix-free implementations of both $\mathbf{G}$ and its
adjoint $\mathbf{G}^H$ required by the posterior sampling step.
We define the three-channel elastic model as the concatenated vector
\begin{equation}
    \mathbf{m} =
    \begin{bmatrix}
        \mathbf{v}_p \\
        \mathbf{v}_s \\
        \boldsymbol{\rho}
    \end{bmatrix}
    \sim p_m(\mathbf{v}_p, \mathbf{v}_s, \boldsymbol{\rho}),
    \label{eq:model}
\end{equation}
where $\mathbf{v}_p, \mathbf{v}_s, \boldsymbol{\rho} \in \mathbb{R}^N$
are the P-wave velocity, S-wave velocity, and density models defined on an
$n_x \times n_z$ spatial grid ($N = n_x n_z$ total grid points),
with $(x, z)$ representing the horizontal and depth dimensions,
respectively.
The corresponding diffusion state at step $t$ is denoted as
\begin{equation}
    \mathbf{m}_t =
    \begin{bmatrix}
        \mathbf{v}_{p_t} \\
        \mathbf{v}_{s_t} \\
        \boldsymbol{\rho}_t
    \end{bmatrix}.
    \label{eq:model_t}
\end{equation}

\subsection{Diffusion Prior over Elastic Parameters}

We employ a Denoising Diffusion Probabilistic Model \citep[DDPM;][]{ho2020denoising}
to learn a prior distribution $p(\mathbf{m})$ over the elastic model space,
providing a flexible, data-driven prior that encodes the statistical
structure of geologically realistic subsurface models \citep{chung2022improving}.
The forward diffusion process progressively corrupts the elastic model
$\mathbf{m}$ by injecting Gaussian noise at each timestep
$t \in \{1, \ldots, T\}$:
\begin{equation}
    \mathbf{m}_t = \sqrt{\alpha_t}\,\mathbf{m} + \sqrt{1 - \alpha_t}\,\boldsymbol{\varepsilon},
    \qquad \boldsymbol{\varepsilon} \sim \mathcal{N}(\mathbf{0}, \mathbf{I}),
    \label{eq:forward_diffusion}
\end{equation}
where $\alpha_t = \prod_{i=1}^{t}(1 - \beta_i)$ is the cumulative noise
schedule, $\{\beta_i\}$ is a predefined variance schedule, and $T$ is the
total number of diffusion timesteps.
The forward process jointly corrupts all three elastic parameter channels,
such that as $t \rightarrow T$ the joint distribution converges to an
isotropic Gaussian in the full elastic parameter space.

\subsection{The reverse Diffusion Process and Score Function}

The reverse diffusion process learns to denoise $\mathbf{m}_t$ back to
$\mathbf{m}_0$ by training a 2D U-Net
$\boldsymbol{\varepsilon}_\theta(\mathbf{m}_t, t)$ to predict the injected
noise across all three channels.
Using Tweedie's formula, an estimate of the clean elastic model at
any noisy state $\mathbf{m}_t$ is obtained as
\begin{equation}
    \hat{\mathbf{m}}_0(\mathbf{m}_t) =
    \frac{\mathbf{m}_t - \sqrt{1 - \alpha_t}\,\boldsymbol{\varepsilon}_\theta(\mathbf{m}_t, t)}{\sqrt{\alpha_t}}.
    \label{eq:tweedie}
\end{equation}
The prior score function, required for posterior sampling, is recovered
from the predicted noise as
\begin{equation}
    \nabla_{\mathbf{m}_t} \log p(\mathbf{m}_t)
    \approx -\frac{\boldsymbol{\varepsilon}_\theta(\mathbf{m}_t, t)}{\sqrt{1 - \alpha_t}}.
    \label{eq:score}
\end{equation}
The unconditional reverse diffusion step, which propagates the model
from state $\mathbf{m}_t$ to $\mathbf{m}_{t-1}$ using only the learned
prior, is given by \citep{ho2020denoising, song2020denoising}:
\begin{equation}
    \mathbf{m}_{t-1} = \sqrt{\alpha_{t-1}}\,\hat{\mathbf{m}}_0(\mathbf{m}_t)
    + \sqrt{1 - \alpha_{t-1} - \sigma_t^2}\,
    \boldsymbol{\varepsilon}_\theta(\mathbf{m}_t, t)
    + \sigma_t\,\boldsymbol{\varepsilon},
    \qquad \boldsymbol{\varepsilon} \sim \mathcal{N}(\mathbf{0}, \mathbf{I}),
    \label{eq:unconditional_update}
\end{equation}
where
\begin{equation}
    \sigma_t = \eta\sqrt{\frac{1 - \alpha_{t-1}}{1 - \alpha_t}}
    \sqrt{1 - \frac{\alpha_t}{\alpha_{t-1}}},
    \label{eq:sigma}
\end{equation}
with $\eta \in [0,1]$ controlling the level of stochasticity:
$\eta = 1$ recovers the original DDPM ancestral sampling
\citep{ho2020denoising}, while $\eta = 0$ yields the deterministic
DDIM sampler \citep{song2020denoising}.
This unconditional update serves as the base sampling step; in
Section~\ref{sec:dps} it is augmented with a data-likelihood
gradient to perform posterior sampling conditioned on the observed
seismic angle-stack data.

\subsection{Diffusion Posterior Sampling (DPS) for AVA Inversion}\label{sec:dps}

To condition the reverse diffusion process on the observed seismic
angle-stack data $\mathbf{d}$, we employ the Diffusion Posterior Sampling
(DPS) framework \citep{chung2022improving}, which decomposes the
posterior score via Bayes' rule as:
\begin{equation}
    \nabla_{\mathbf{m}_t} \log p(\mathbf{m}_t \mid \mathbf{d})
    = \underbrace{\nabla_{\mathbf{m}_t} \log p(\mathbf{m}_t)}_{\text{prior score (Eq.~\ref{eq:score})}}
    + \nabla_{\mathbf{m}_t} \log p(\mathbf{d} \mid \mathbf{m}_t),
    \label{eq:posterior_score}
\end{equation}
where the first term is the unconditional prior score from the pretrained
diffusion model (Equation~\ref{eq:score}), and the second term introduces
a data-consistency correction that steers samples toward models consistent
with the observed angle-stack data.

The likelihood term $p(\mathbf{d} \mid \mathbf{m}_t)$ is intractable
directly in the noisy latent space. DPS approximates it by substituting
the Tweedie denoised estimate $\hat{\mathbf{m}}_0(\mathbf{m}_t)$ from
Equation~\ref{eq:tweedie} into the forward model
(Equation~\ref{eq:forward_matrix}), yielding the likelihood gradient
with respect to $\mathbf{m}_t$ as
\begin{equation}
    \nabla_{\mathbf{m}_t} \log p(\mathbf{d} \mid \mathbf{m}_t)
    \approx \frac{1}{\sigma_d^2}
    \left(\nabla_{\mathbf{m}_t} \hat{\mathbf{m}}_0(\mathbf{m}_t)\right)^T
    \mathbf{G}^H \left(\mathbf{d} - \mathbf{G}\hat{\mathbf{m}}_0(\mathbf{m}_t)\right),
    \label{eq:likelihood_grad}
\end{equation}
where $\sigma_d^2$ is the data noise variance (introduced in
Equation~\ref{eq:forward_conv}) and $\mathbf{G}^H$ is the adjoint of the
AVA forward operator defined in Equation~\ref{eq:forward_matrix},
which maps the data residual back into the model space
\citep{ravasi2020pylops}.
The corresponding DPS gradient descent direction in model space is
\begin{equation}
    \mathbf{g}_t := -\nabla_{\mathbf{m}_t} \log p(\mathbf{d} \mid \mathbf{m}_t)
    \approx -\frac{1}{\sigma_d^2}
    \left(\nabla_{\mathbf{m}_t} \hat{\mathbf{m}}_0(\mathbf{m}_t)\right)^T
    \mathbf{G}^H \left(\mathbf{d} - \mathbf{G}\hat{\mathbf{m}}_0(\mathbf{m}_t)\right),
    \label{eq:dps_grad}
\end{equation}
which represents a steepest-descent update honoring data consistency
in the elastic model space.
Combining the prior score (Equation~\ref{eq:posterior_score}) and the
likelihood gradient $\mathbf{g}_t$ (Equation~\ref{eq:dps_grad}),
the full guided reverse diffusion update at each step $t$ extends
the unconditional update of Equation~\ref{eq:unconditional_update} as
\begin{equation}
    \mathbf{m}_{t-1} = \sqrt{\alpha_{t-1}}\,\hat{\mathbf{m}}_0(\mathbf{m}_t)
    + \sqrt{1 - \alpha_{t-1} - \sigma_t^2}\,
    \boldsymbol{\varepsilon}_\theta(\mathbf{m}_t, t)
    + \sigma_t \boldsymbol{\varepsilon}
    - \mu\, \mathbf{g}_t,
    \label{eq:update}
\end{equation}
where $\sigma_t$ is defined in Equation~\ref{eq:sigma}, and $\mu$ is a
scalar weighting factor that controls the guidance strength i.e., the relative influence of the data likelihood correction relative
to the prior. Setting $\mu = 0$ recovers the unconditional prior sampling
of Equation~\ref{eq:unconditional_update}, while larger values enforce
stronger data consistency at each step.

\subsection{Model training}\label{sec:model_train_init}
\begin{figure}
\centering
\includegraphics[scale=0.17]{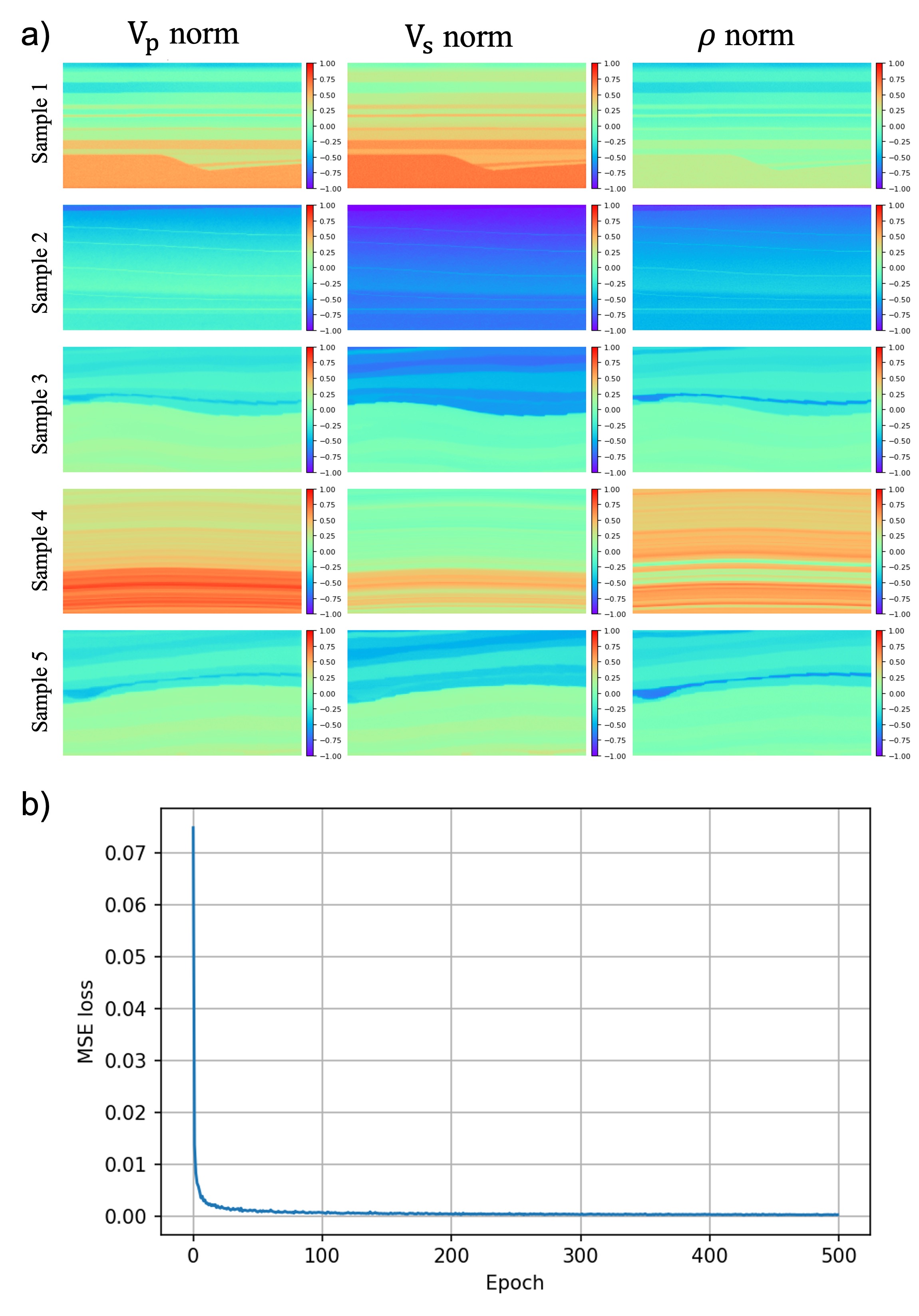} 
\caption{DDPM training results. a) Five representative unconditional samples generated by the trained diffusion model, showing the three normalized elastic parameters i.e., $V_p$ norm, $V_s$ norm, and $\rho$ norm and each mapped to the range [$-$1, 1] via global per-channel min–max normalization. b) MSE training loss curve as a function of epoch.}

\label{figunc} 
\end{figure}

A DDPM with a 2D U-Net backbone is pretrained to learn a prior
distribution from high-resolution elastic models by approximating
the score function $\nabla_{\mathbf{m}_t} \log p(\mathbf{m}_t)$
(Equation~\ref{eq:score}).
The training dataset consists of 5000 high-resolution elastic models
drawn from the SEAM Arid, Overthrust, and Marmousi2 benchmark models,
supplemented by 2D models generated from field well-log profiles.
All models are normalized to the range $[-1, 1]$ using global per-channel
min-max normalization before being input to the DDPM.
All the elastic models of size $128 \times 256$ are used for training 
and augmented through elastic transformations that introduce geological
structural complexity such as folds and faults, improving the
generalization of the learned prior to out-of-distribution subsurface
scenarios. The network is trained for 500 epochs using $T = 1000$ diffusion time-steps with the Adam optimizer \citep{kingma2014adam}. Figure \ref{figunc}a shows the unconditional samples which confirms that the DDPM has successfully learned a geologically meaningful prior over the joint elastic parameter distribution. The generated $V_p$, $V_s$, and $\rho$ models exhibit consistent horizontal layering, smooth lateral continuity, and physically consistent structural features. Additionally, Figure \ref{figunc}b illustrates training curve plot wherein a rapid convergence of the plot is observed with the MSE loss dropping sharply within the first 50 epochs and plateauing near zero by epoch 100 and remains stable through epoch 500.

\subsection{Uncertainty quantification}
\subsubsection{Uncertainty Quantification with Diffusion Posterior Sampling}\label{sec:dps_uq}
Uncertainty in the inverted elastic parameters is quantified using
diffusion posterior sampling (DPS). The method generates samples from
the posterior distribution by guiding the reverse diffusion process
with the gradient of the data misfit, as described in
Section~\ref{sec:dps}.\\
Starting from Gaussian noise, multiple realizations of the model
$\mathbf{m}$ are generated by solving the stochastic differential
equation associated with the reverse diffusion process. In this work, the stochasticity of the reverse process is fully retained
by fixing the stochasticity level of the diffusion transitions to unity ($\eta$=1, in equation \ref{eq:sigma}), corresponding to a DDPM formulation. This ensures that the generated samples fully explore the posterior variability, rather than collapsing to deterministic trajectories. Each realization is conditioned on the observed data through the likelihood term, resulting in a set of samples approximately distributed according to the posterior $p(\mathbf{m} \mid \mathbf{d})$. Given $N_s$ posterior samples $\{\mathbf{m}^{(k)}\}_{k=1}^{N_s}$, the posterior mean and standard deviation are estimated as

\begin{equation}
\bar{\mathbf{m}} =
\frac{1}{N_s} \sum_{k=1}^{N_s} \mathbf{m}^{(k)},
\end{equation}
\begin{equation}
\boldsymbol{\sigma}_{\mathrm{DPS}} =
\sqrt{
\frac{1}{N_s - 1}
\sum_{k=1}^{N_s}
\left(
\mathbf{m}^{(k)} - \bar{\mathbf{m}}
\right)^2
}.
\end{equation}
The resulting standard deviation provides a spatially resolved measure
of uncertainty that reflects both the data constraints and the
 non-Gaussian prior learned by the diffusion model. 

\subsubsection{Analytical Bayesian Uncertainty Quantification}\label{sec:ana_uq}

To benchmark the uncertainty quantification obtained from diffusion posterior sampling, we additionally consider an analytical Bayesian inversion framework, similar to \cite{buland2003bayesian}.
The observed data are modeled as equation \ref{eq:forward_matrix}.
A Gaussian prior is imposed on the model,
\begin{equation}
\mathbf{m} \sim \mathcal{N}(\boldsymbol{\mu}, \boldsymbol{\Sigma}_m),
\end{equation}
where $\boldsymbol{\mu}$ corresponds to the smooth initial model. The prior covariance
$\boldsymbol{\Sigma}_m$ is defined implicitly through a quadratic
regularization operator. In particular, the corresponding precision
matrix is expressed as
\begin{equation}
\boldsymbol{\Sigma}_m^{-1}
=
\lambda_I \mathbf{I}
+
\lambda_{lap} \mathbf{D}^T \mathbf{D},
\end{equation}
where $\mathbf{D}$ denotes a second-order derivative operator acting
along the depth axis. This formulation enforces proximity to the prior
model and promotes vertical smoothness of the elastic parameters. The analytical form of the posterior distribution is obtained by assuming a linear forward model with Gaussian prior and noise distributions, leading to a Gaussian posterior. This allows for closed-form expressions of both the posterior mean and covariance.
Under these assumptions, the posterior distribution is Gaussian,
\begin{equation}
p(\mathbf{m} \mid \mathbf{d}) \sim
\mathcal{N}(\boldsymbol{\mu}_{post}, \boldsymbol{\Sigma}_{post}),
\end{equation}
with mean and covariance given by
\begin{equation}
\boldsymbol{\mu}_{post}
=
\boldsymbol{\mu}
+
\boldsymbol{\Sigma}_m \mathbf{G}^T
\left(
\mathbf{G} \boldsymbol{\Sigma}_m \mathbf{G}^T
+ \sigma_d^2 \mathbf{I}
\right)^{-1}
\left(
\mathbf{d} - \mathbf{G}\boldsymbol{\mu}
\right),
\end{equation}
\begin{equation}
\boldsymbol{\Sigma}_{post}
=
\boldsymbol{\Sigma}_m
-
\boldsymbol{\Sigma}_m \mathbf{G}^T
\left(
\mathbf{G} \boldsymbol{\Sigma}_m \mathbf{G}^T
+ \sigma_d^2 \mathbf{I}
\right)^{-1}
\mathbf{G} \boldsymbol{\Sigma}_m.
\end{equation}

To ensure computational tractability, the inversion is performed
independently for each spatial location, assuming a locally
one-dimensional Earth model for each angle-stack gather. Under
this assumption, the posterior distribution is evaluated tracewise, yielding a local posterior covariance defined over the depth samples and elastic parameters at each trace. This significantly reduces the dimensionality of the problem and enables efficient evaluation of the posterior uncertainty. This results in a posterior covariance defined per trace, rather than over the full spatial domain. Uncertainty is quantified through the posterior standard deviation, computed as the square root of the diagonal elements of $\boldsymbol{\Sigma}_{post}$. \label{sec:dps_uq}

\section{Results}

\subsection{Synthetic example: Otway model (out-of-distribution example)}
\begin{figure} 
\centering
\includegraphics[scale=0.28]{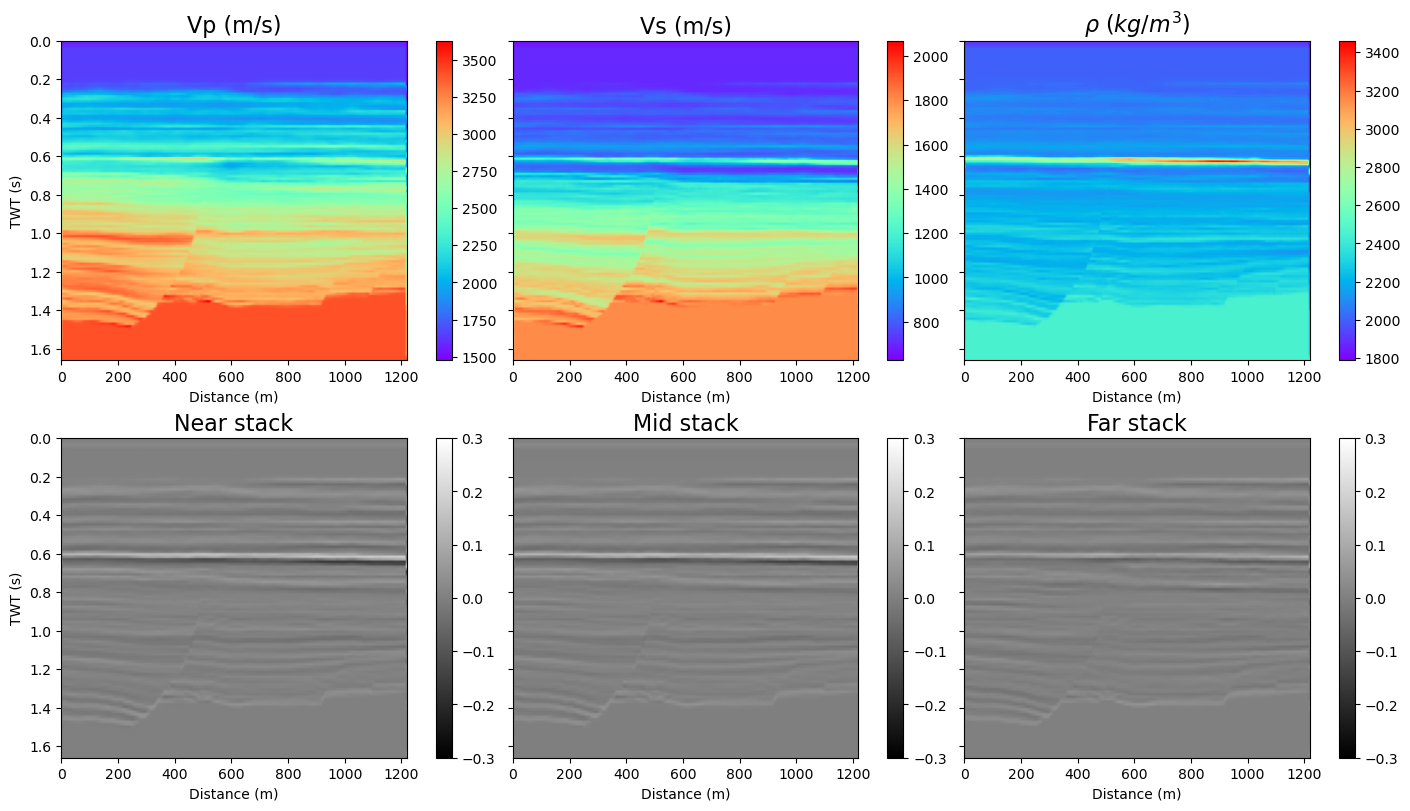} 
\caption{True elastic models of Otway data (in the top row) and synthetically generated three angle-stack seismic section (in the bottom row). } 
\label{fig1} 
\end{figure}

We first evaluate the effectiveness of the proposed guidance-based diffusion framework on a synthetic 2D Otway elastic model. To incorporate guidance during the reverse diffusion process using data misfit (Eq. \ref{eq:dps_grad}), we generated three angle-stacks with different angle ranges: near-stack (3°-10°), mid-stack (10°–20°) and far-stack (20°–30°). To construct the angle-stacks, we first computed reflectivity from the true elastic models using the Aki–Richards linearized approximation (Eq. \ref{eq:aki_richards}) and then convolved each stack’s reflectivity with a Ricker wavelet of peak frequency 150 Hz. The intuition to choose a high peak frequency is to better capture the fine-layer features present in the Otway model. Figure \ref{fig1} illustrates the true elastic models and the synthesized three angle-stacks. In addition, for the inversion we have added $3\%$ Gaussian noise to the generated angle-stacks.

\begin{figure}
\centering
\includegraphics[scale=0.28]{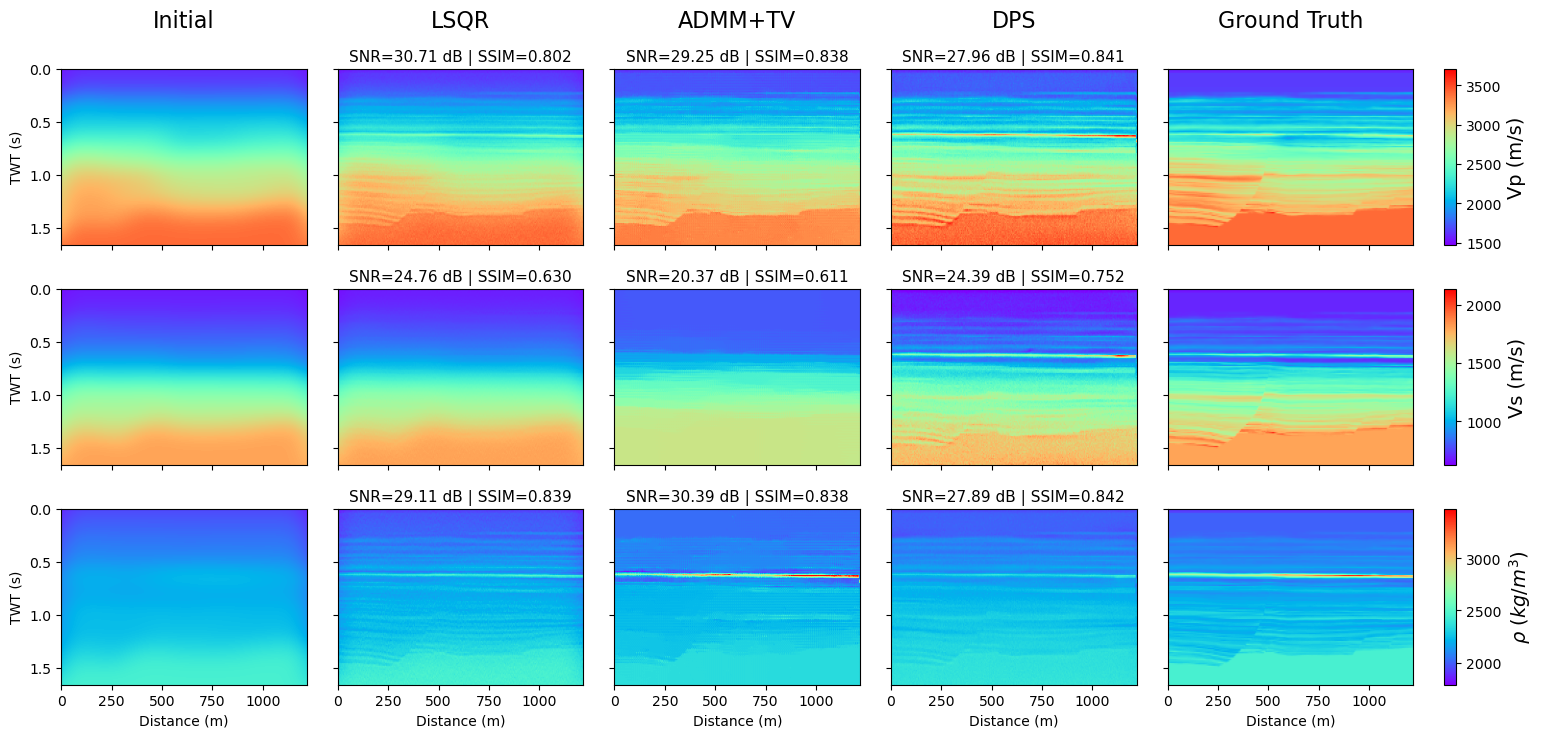} 
\caption{Inverted elastic parameters for the Otway model (out-of-distribution test). The first column shows the initial low-frequency elastic models. The second column presents the inverted elastic parameters obtained with a conventional LSQR. The third column displays the
recovered elastic parameters using ADMM with total variation regularization. The fourth column shows the recovered elastic parameters generated by the proposed guidance-based diffusion approach (DPS guidance). The final column presents the true elastic models of the Otway synthetic model.} 
\label{fig2} 
\end{figure}

\begin{figure}
\centering
\includegraphics[scale=0.3]{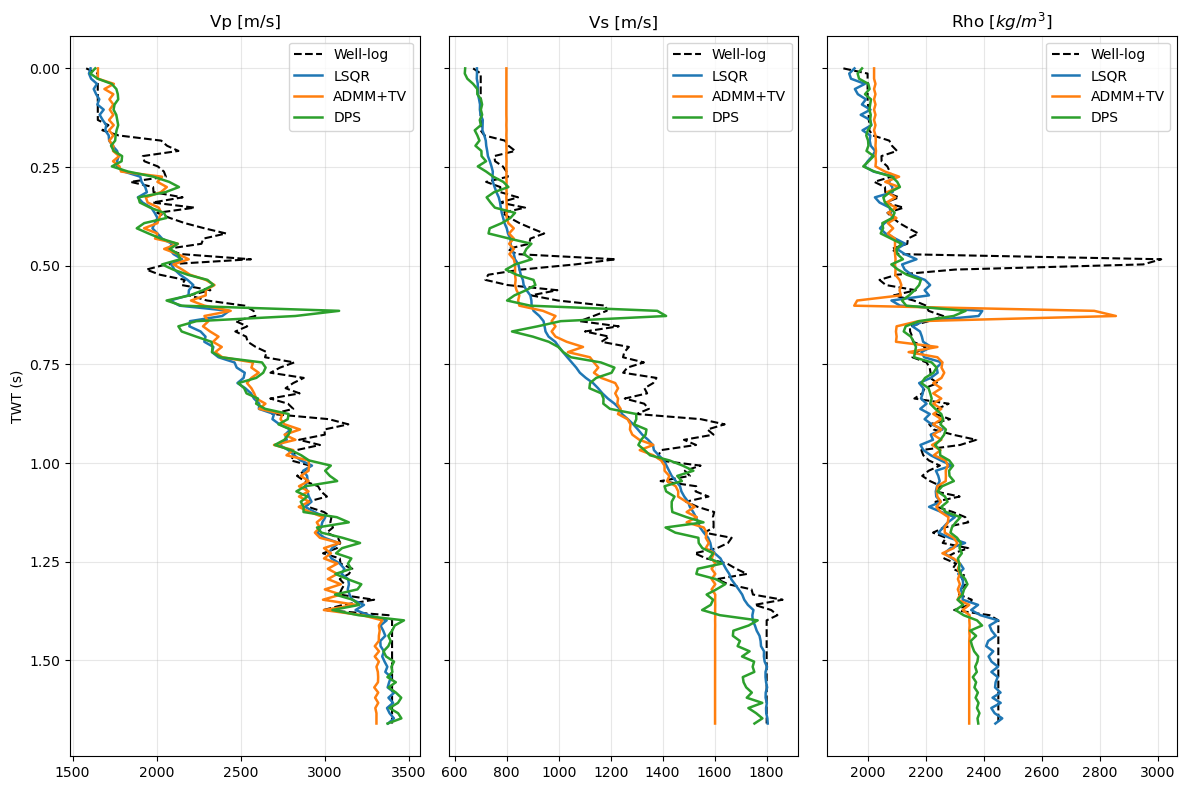} 
\caption{1D comparison of inverted elastic parameters for the Otway model (out-of-distribution test) at a representative trace location. Each panel displays the vertical profile of $V_p$(left), $V_s$ (middle), and density $\rho$ (right) as a function of two-way traveltime (TWT). The dashed black curve represents the ground-truth trace, while the blue, orange, and green curves correspond to the elastic parameters recovered by LSQR, ADMM with total variation regularization, and the proposed guidance-based diffusion approach (DPS guidance), respectively.} 
\label{fig3} 
\end{figure}

Figure \ref{fig2} compares inverted elastic models for the Otway synthetic model, recovered by least-squares QR (LSQR), ADMM with total-variation regularization (ADMM+TV), and Diffusion Posterior Sampling (DPS), alongside the initial model and ground truth. $V_p$ recovery is strong across all three methods. LSQR achieved the highest SNR (30.71 dB, SSIM = 0.802), followed by ADMM+TV (29.25 dB, SSIM = 0.838) and DPS (27.96 dB, SSIM = 0.841). In particular, DPS yielded the best structural recovery, as reflected by its superior SSIM, suggesting that diffusion prior better captures fine-scale layering and lateral continuity in $V_p$. $V_s$ proved to be the most challenging parameter, with all methods producing lower SNR and SSIM values compared to $V_p$ and $\rho$. LSQR attained the highest SNR (24.76 dB, SSIM = 0.630), while DPS achieved the best structural similarity (SSIM = 0.752), demonstrating the diffusion prior’s ability to recover lateral continuity in $V_s$. ADMM+TV performed worst on both metrics (SNR = 20.37 dB, SSIM = 0.611), likely due to over-smoothing introduced by TV regularization for this more heterogeneous parameter.$\rho$ recovery was consistently strong across methods. ADMM+TV achieved the best SNR (30.39 dB, SSIM = 0.838), closely followed by LSQR (SNR = 29.11 dB, SSIM = 0.839), which attained the highest structural similarity. DPS remained competitive (SNR = 27.89 dB, SSIM = 0.842) and produced the highest SSIM for $\rho$, preserving the sharp impedance contrast at the shallow reflector, consistent with the visually well-defined reflector in the DPS density panel. In addition, Figure \ref{fig3} shows a 1D trace comparison at the location 612 m in Figure \ref{fig2}, comparing the three inversion methods against the ground truth. For $V_p$, all methods broadly recovered the true trend over the full TWT range (~0–1.6 s). LSQR and ADMM+TV exhibit larger high-frequency wiggles and occasional offsets, whereas DPS is comparatively smoother. For $V_s$, ADMM+TV captures local variations well, but ADMM+TV shows an anomalous vertical spike around TWT ~1.4–1.6 s, possibly indicating numerical instability or ill-conditioning in the TV-regularized inversion. LSQR and DPS remain stable and consistent overall, with DPS recovering richer structural detail and LSQR producing a smoother estimate. $\rho$ is reasonably well recovered by all three methods; however, ADMM+TV produces an overestimated sharp peak at ~0.62 s. Furthermore, none of the methods is able to recover the true sharp density spike at TWT ~0.5 s.

Overall, DPS showcases a balanced recovery across all three elastic parameters, with particular advantage in Vs structural accuracy. Although LSQR and ADMM+TV slightly surpass DPS in peak SNR for $V_p$ and density, DPS consistently produces structurally consistent results, which is an important quality for AVO/AVA interpretation workflows.

\subsection{Field data example: Poseidon dataset}
\begin{figure}
\centering
\includegraphics[scale=0.15]{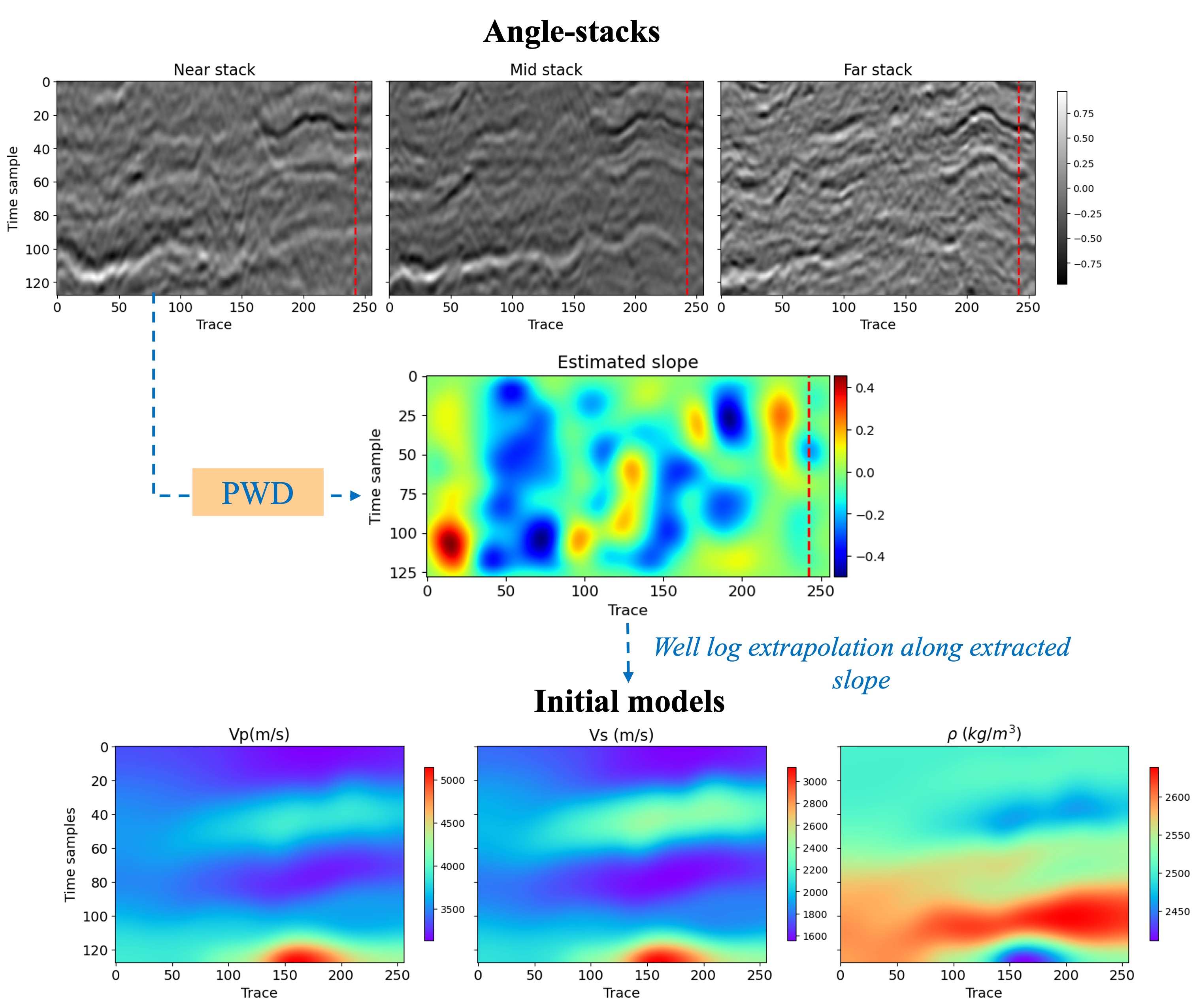} 
\caption{Angle-stack data, estimated slope, and initial models used in the inversion workflow for Field data implementation. Top: Near, mid, and far angle stacks. Middle: Local seismic slope field estimated from the near-angle stack using plane-wave destruction (PWD) solver. Bottom: Initial smooth elastic models for $V_p$, $V_s$, and $\rho$.
Red dashed line indicating the location of the well.}
\label{fig4} 
\end{figure}

In this section, we demonstrate the application of the proposed
framework on the field data from Poseidon dataset, NW-
Shelf in the Browse basin, Australia. We utilize three angle-
stacks and four wells, which are publicly available. The three
angle-stacks have the following: near-stack (6° $-$ 18°),
mid-stack (18° $-$ 30°) and far-stack (30° $-$ 42°). The
statistical wavelet is extracted from each of the stacks and then utilized to generate the synthetic stacks using linearized Zoeppritz operator at the time of reverse diffusion process for guidance. Obtaining a geologically consistent initial model is critical for field data given the structural and stratigraphic complexity inherent in such datasets \citep{yang2025hctnet}. To obtain a structurally consistent initial model, we propagate well-log measurements laterally along estimated structural dips using plane-wave destruction (PWD) filters \citep{fomel2002applications} derived from the near-angle stack. This is achieved by solving a structurally preconditioned well-matching Tikhonov inversion problem, which enforces geological consistency between the initial model and the observed structural trends in the seismic data \citep{brandolin2026velocity}. Subsequently, a Gaussian smoothing filter is applied to reconstruct a geologically plausible low-frequency elastic model. Figure \ref{fig4} illustrates the field angle-stack data along with the estimated slope using the PWD solver. The slope information is utilized to extrapolate well-logs along the 2D section. Afterward, the smoothed version of the extrapolated well-log sections was used as initial model for the inference, which initializes the reverse diffusion chain.  

\begin{figure}
\centering
\includegraphics[scale=0.25]{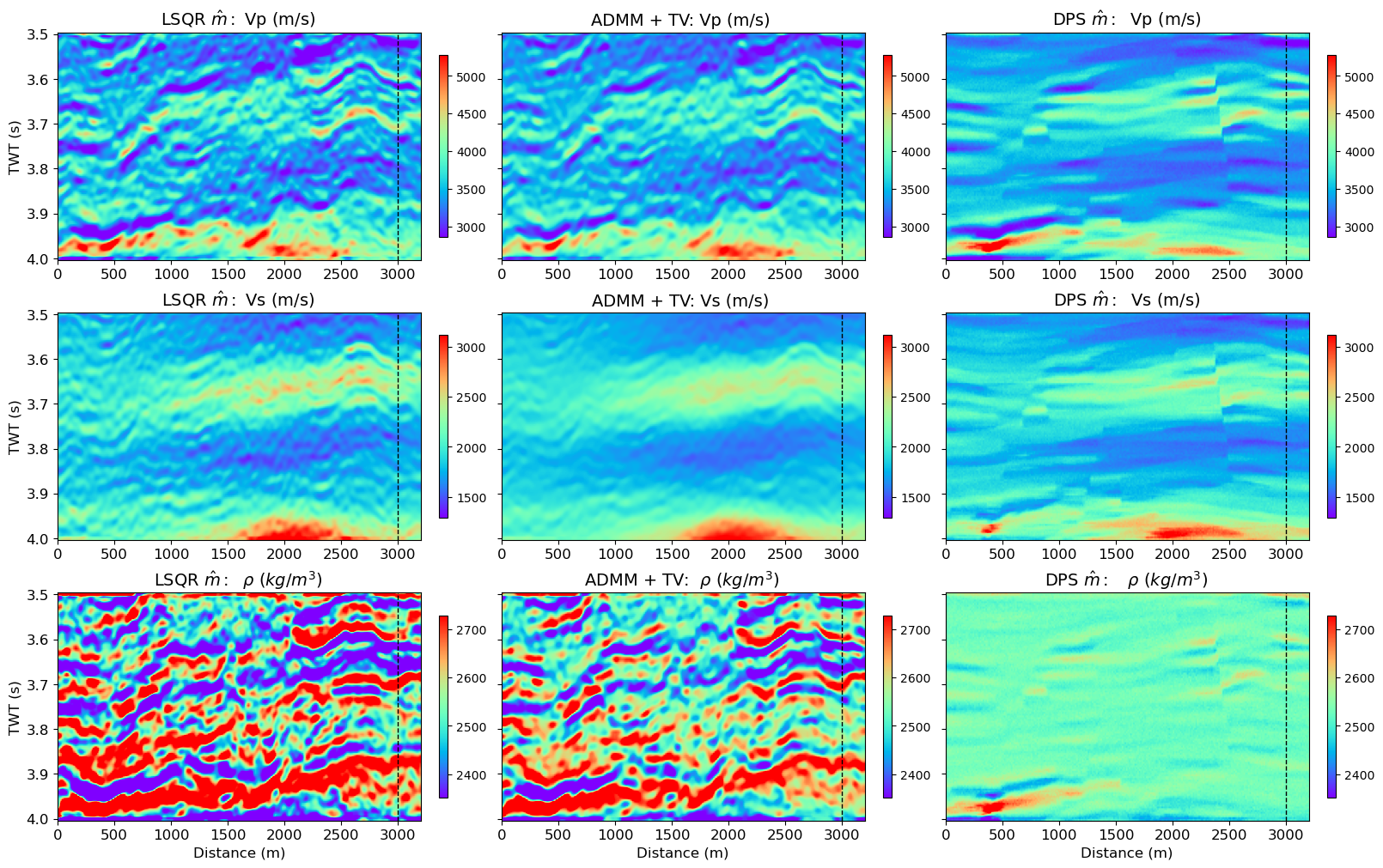} 
\caption{Inverted elastic parameters for the Poseidon field (field data). Rows: $V_p$ (top), $V_s$ (middle), and $\rho$ (bottom). The first column presents the inverted elastic parameters obtained with conventional LSQR. The second column displays the recovered elastic parameters using ADMM with total-variation regularization (ADMM+TV). The third column shows the recovered elastic parameters produced by the proposed guidance-based diffusion approach (DPS). The dashed vertical line marks the well location shown in well-comparison plots.} 
\label{fig5} 
\end{figure}

Figure \ref{fig5} presents the inverted elastic parameter sections $V_p$, $V_s$, and $\rho$ recovered from the field dataset using LSQR, ADMM+TV, and the proposed DPS approach, alongside the initial low-frequency models. The dashed vertical line at approximately 3000 m marks the well location used as a reference for validation. The inversion results are shown for the deep gas-reservoir interval in the field. The conventional LSQR inversion substantially increases spatial variability but introduces strong noise and streaky artifacts, especially in the deeper part of the section, which can obscure true geological features. The ADMM+TV solution effectively suppresses much of the noise and yields more laterally continuous layers, but at the cost of over-regularizing sharp boundaries and smearing fine-scale contrasts. In contrast, the DPS-guided inversion produces elastic models that retain coherent, continuous reflectors while also preserving sharper vertical and lateral contrasts, leading to a visually more interpretable result that aligns better with expected geological structure from the Poseidon field. In addition to gain more level of confidence we also compared the inverted results with the well (shown as black dashed line in Figure \ref{fig5}). 

\begin{figure}
\centering
\includegraphics[scale=0.29]{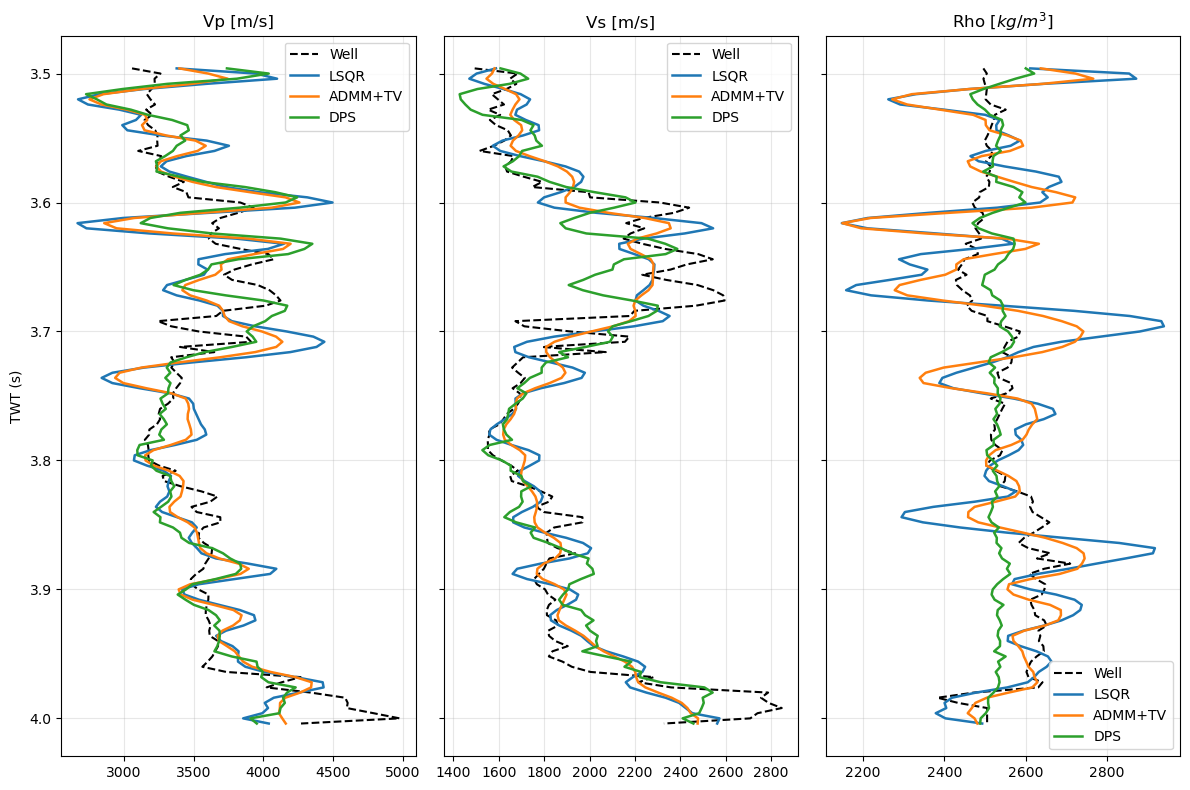} 
\caption{Comparison between true well-log profiles and inverted elastic parameters. Left: $V_p$, middle: $V_s$, and right: $\rho$.Each panel shows the reference well log (black dashed) and inversion results obtained with LSQR (orange), ADMM+TV (green), and DPS (blue).}

\label{fig6} 
\end{figure}

Figure \ref{fig6} presents a 1D comparison with well-logs. All three inversion methods recover the broad $V_p$ trend observed in the well-log. Although they struggle to recover the match with the $V_p$ log especially for the top part starting from ~3.5 to 3.7 s. However, DPS inversion curve tends to be closer to the well-log as compared to LSQR and ADMM+TV. After 3.7 s, LSQR and ADMM+TV show peaks of over- and underestimation of the $V_p$ whereas DPS is in more agreement with the $V_p$ log. All three inversion methods recover Vs reasonably well overall. Between 3.60–3.70 s, the DPS trace departs from the well log showing a modest underestimation and a smoother, shifted response relative to the $V_s$ log where LSQR and ADMM+TV more closely follow the trend. However, after ~3.70 s DPS realigns with the well and tracks the $V_s$ log variations up to ~4.00 s with good agreement. For density, the conventional inversions systematically overestimate values and exhibit exaggerated high-frequency variations, producing larger deviations from the density log. ADMM+TV reduces some extremes compared with LSQR but still shows notable offsets at several intervals. By contrast, the DPS-guided inversion yields density values that lie within the expected range and recovers a smooth, geologically consistent trend that aligns best with the well. Nevertheless, DPS smooths out fine-scale density fluctuations present in the log and underestimates a few sharp jumps. 

\subsection{Posterior uncertainty quantification}

\subsubsection{Synthetic Otway Data}
\begin{figure}
\centering
\includegraphics[scale=0.3]{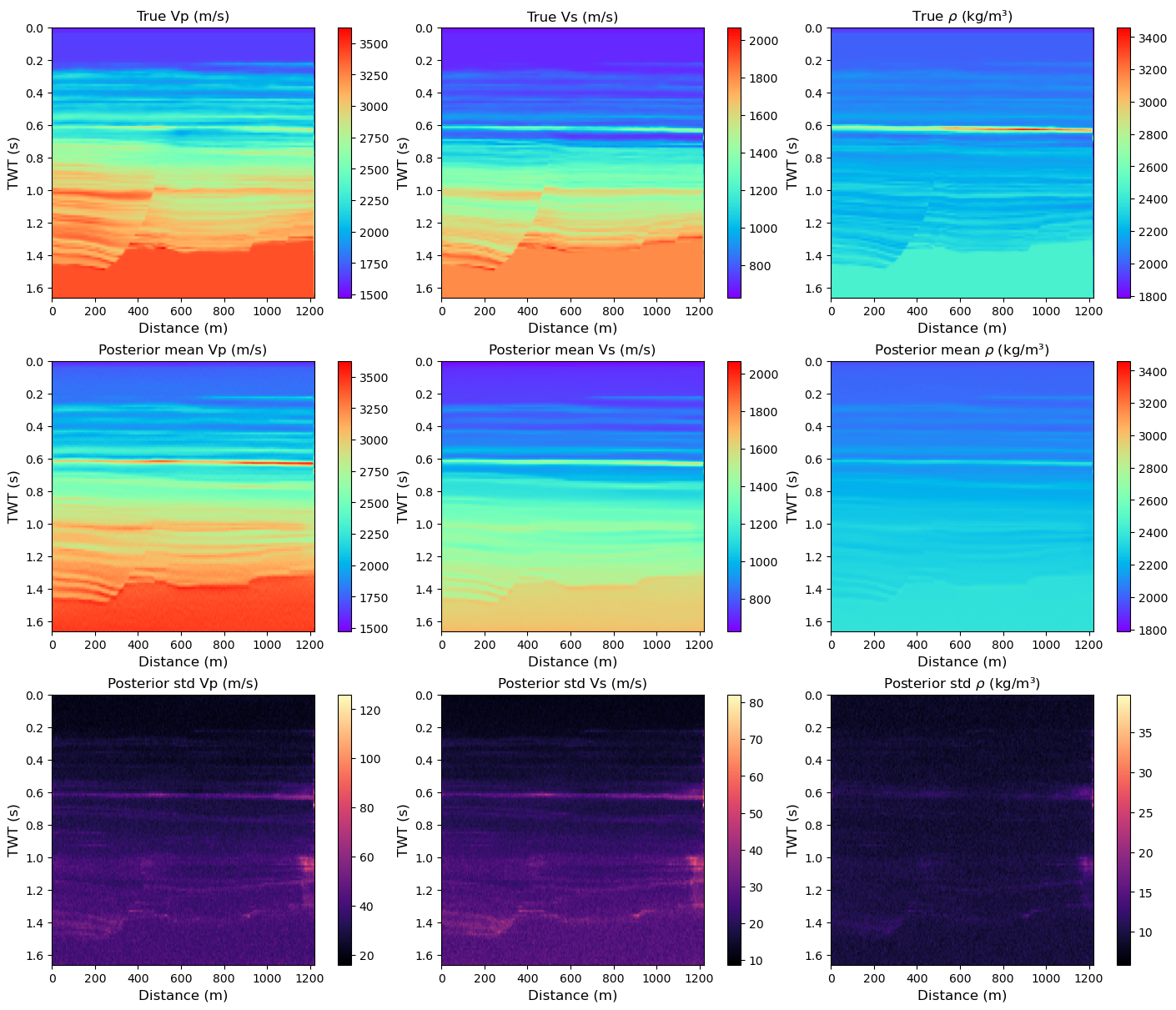} 
\caption{Bayesian seismic inversion result using guidance-based DPS. (Top row) Otway elastic models of $V_p$, $V_s$ and $\rho$, shown as a function of two-way travel time (TWT,s). (middle row) Posterior mean estimates of $V_p$, $V_s$ and $\rho$ recovered by the proposed method. (Bottom row) Posterior standard deviation of $V_p$, $V_s$ and $\rho$} 

\label{fig7} 
\end{figure}

To quantify uncertainty in the inverted elastic parameters using the proposed framework, we generated 100 independent realizations from the posterior distribution. The mean of the generated realizations serves as our prediction and standard deviation of these realizations is used as our measure of uncertainty, as described in Section~\ref{sec:dps_uq}. Figure \ref{fig7} shows results of the Bayesian seismic inversion via guidance-based DPS. The Posterior mean demonstrates close agreement with the true models in terms of large-scale structure, layer geometries, and parameter magnitudes. The posterior standard deviation of $V_p$, $V_s$, and $\rho$, quantifies the spatial distribution of uncertainty. Elevated uncertainty is concentrated near reflector interfaces and in laterally heterogeneous regions (particularly the fault structure visible below ~1.2 s TWT and to the left of ~600 m lateral distance), while well-constrained regions deeper in the section and far from structural complexity exhibit comparatively low uncertainty.

\begin{figure}
\centering
\includegraphics[scale=0.3]{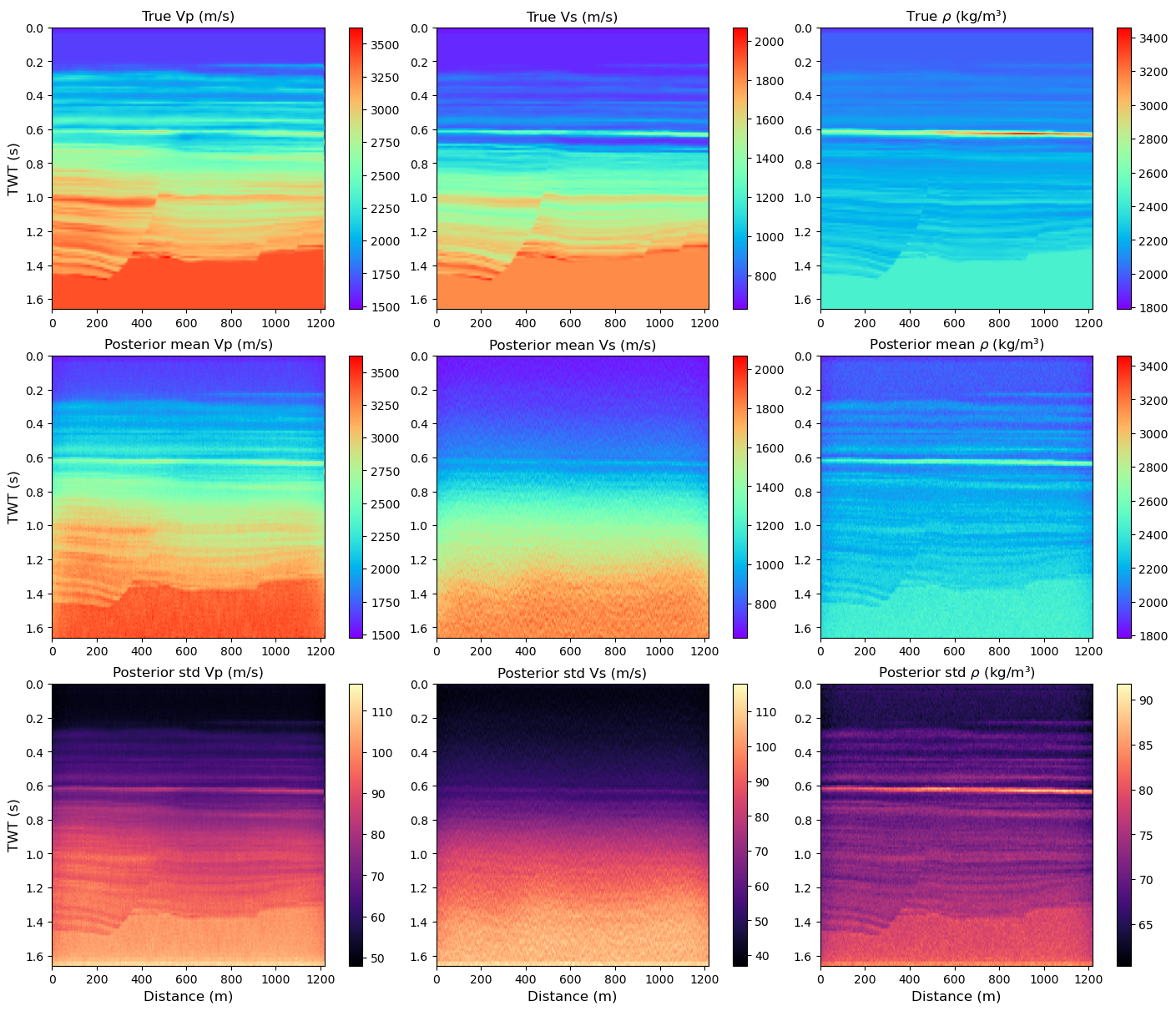} 
\caption{Analytical Bayesian uncertainty quantification results for the Otway synthetic dataset. (Top row) True elastic models of $V_P$ (left), $V_S$ (middle), and $\rho$ (right) used as ground truth. (Middle row) Posterior mean estimates of $V_P$, $V_S$, and $\rho$ recovered by the closed-form analytical Bayesian inversion. (Bottom row) Posterior standard deviation of $V_P$, $V_S$, and $\rho$, quantifying the spatially 
varying inversion uncertainty.} 

\label{fig8} 
\end{figure}

The analytical Bayesian inversion results, derived using the closed-form framework of \citet{buland2003bayesian}, are presented in Figure~\ref{fig8} for comparison. The true models reveal a well-stratified subsurface with a prominent high-density interface at $\approx$0.6~s TWT and a structurally complex fault zone below 1.0~s TWT. The posterior mean recovers the large-scale stratigraphy and parameter magnitudes with good correlation with the true models for all three parameters, most accurately for $V_P$ and with smoothing for $V_S$ and $\rho$, particularly in the laterally heterogeneous fault region where the Gaussian prior regularization suppresses sharp boundary recovery. The posterior standard deviation maps reveal a physically meaningful, spatially structured uncertainty field: $V_P$ uncertainty ranges from $\sim$50 to 120~m/s, with the lowest values in the shallow, well-illuminated section and elevated values at reflector interfaces and in the deep fault zone; $V_S$ uncertainty spans $\sim$40 to 110~m/s in a more spatially diffuse pattern, reflecting the inherently weaker constraint on $V_s$ from angle-stack data; and $\rho$ uncertainty occupies the narrow range of $\sim$65 to 90~kg/m\textsuperscript{3}, with a localized minimum co-located with the high-density interface at $\sim$0.6~s TWT where the strong acoustic contrast provides unusually tight Bayesian constraint. Across all three parameters, uncertainty increases monotonically with depth, consistent with the attenuation of seismic resolution and signal-to-noise ratio, and is in agreement with regions of structural complexity, confirming that the analytical posterior covariance provides geologically interpretable, calibrated uncertainty estimates. These results serve as the exact Gaussian posterior reference against which the ensemble-based uncertainty of the DPS approach is benchmarked, with the comparison revealing that DPS uncertainty bands are systematically overconfident wherein the data-misfit likelihood gradient collapses the diffusion ensemble around low-misfit solutions and the learned prior further concentrates samples toward high-probability training regions, yielding posterior spreads that are narrower than the true posterior covariance across all three elastic parameters throughout the model.

\begin{figure}
\centering
\includegraphics[scale=0.31]{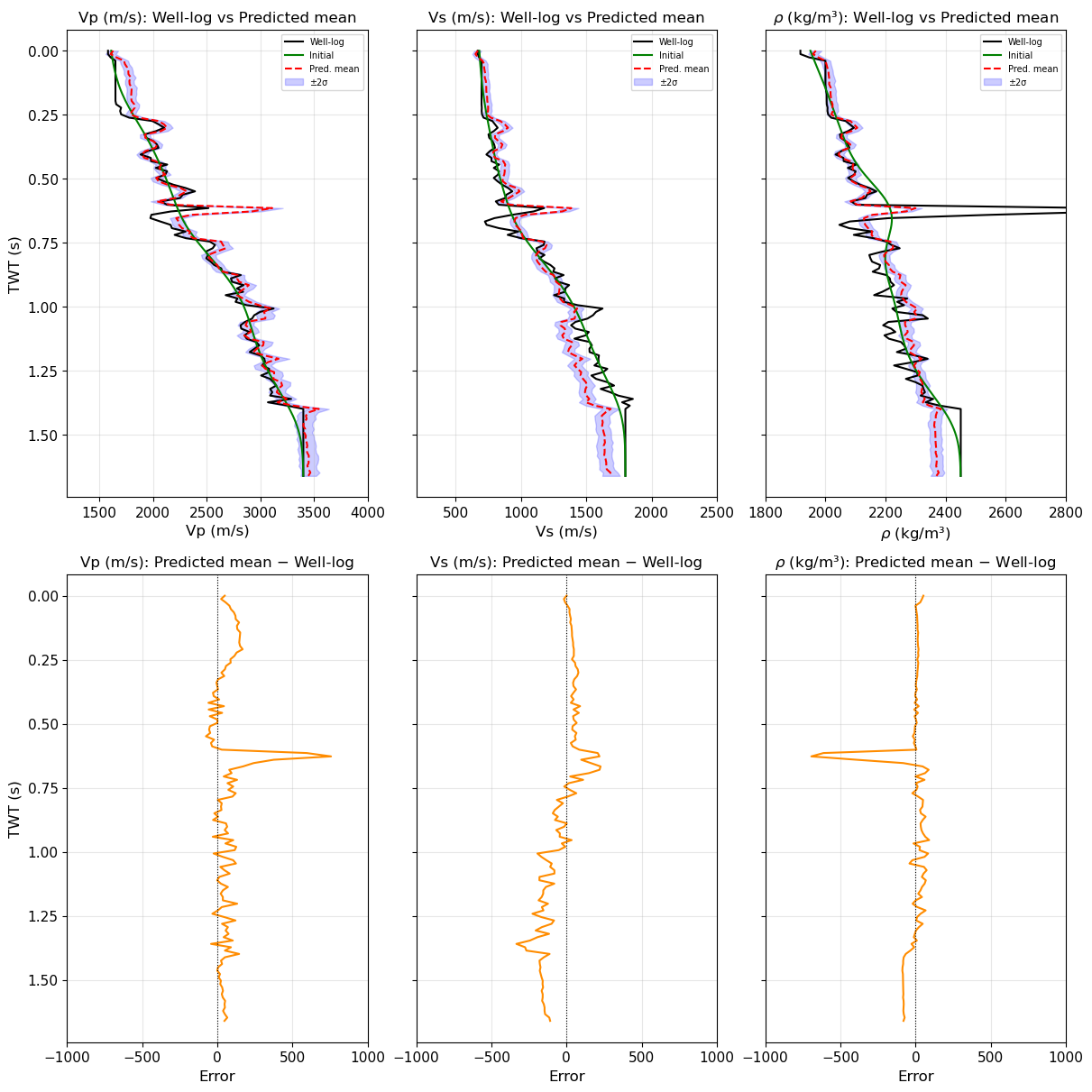} 
\caption{1D comparison of DPS-derived posterior uncertainty at trace position 612 m of the 2D Otway synthetic model. (Top row) profiles of $V_p$, $V_s$ and  $\rho$ along with true log (black), initial smooth model (green), predicted mean (red dashed) and $\pm2\sigma$ uncertainty envelope (blue shading). (Bottom row) vertical error profiles (predicted mean $-$ true log) for $V_p$, $V_s$ and $\rho$ with a dash zero-error reference line; horizontal axes show error magnitude in the same units as each property.} 
\label{fig9} 
\end{figure}

\begin{figure}
\centering
\includegraphics[scale=0.31]{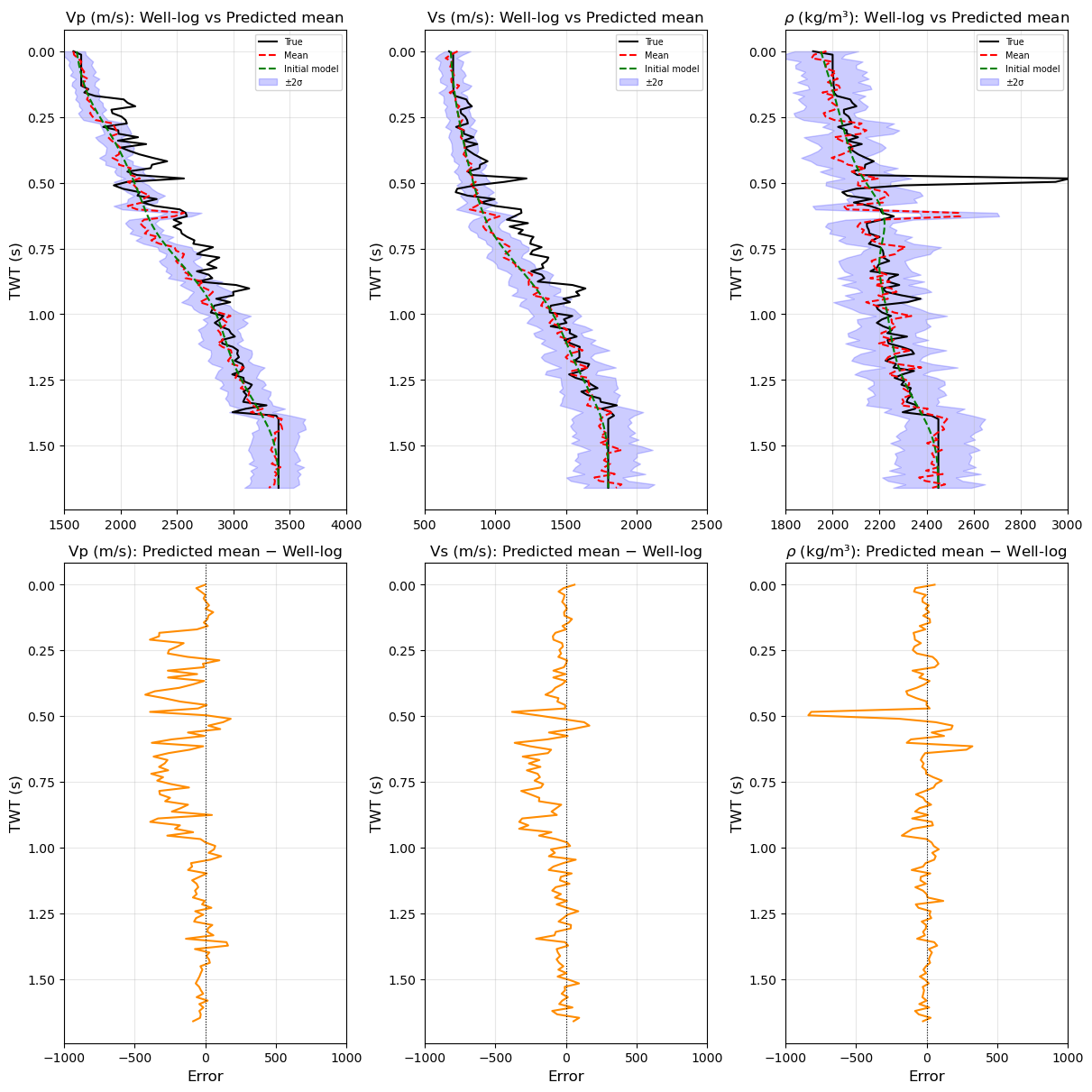} 
\caption{1D comparison of analytical Bayesian posterior uncertainty at trace position 612 m of the 2D Otway synthetic model. (Top row) Vertical profiles of $V_P$ (left), $V_S$ (middle), and $\rho$ (right) as a function of two-way travel time (TWT), showing the true elastic log (black solid), initial smooth model (green dashed), posterior mean (red dashed), and the exact $\pm2\sigma$ uncertainty envelope (blue shading) derived from the closed-form posterior covariance. (Bottom row) Pointwise error profiles (posterior mean $-$ true log) for each elastic parameter.} 
\label{fig10} 
\end{figure}

Consistent with the observations from the 2D posterior maps, the 1D trace comparisons at position 612 m corroborate the same behaviour. For the DPS-based inversion (Figure~\ref{fig9}), the posterior mean recovers the broad elastic trend for all three elastic parameters, yet the $\pm2\sigma$ envelope remains narrow and fails to bracket the true $V_S$ and $\rho$ logs over substantial depth intervals, with a notable overestimation of $V_P$ near 0.6~s TWT consistent with the elevated error patterns observed in the 2D maps. For the analytical Bayesian solution (Figure~\ref{fig10}), the same localised overestimation at 0.6~s TWT is present; however, the $\pm2\sigma$ envelope is markedly wider and better calibrated, representing the true log across the full TWT range for all three elastic
parameters, directly reflecting the broader and geologically structured uncertainty captured by the exact posterior covariance in the 2D standard deviation maps.

\subsubsection{Poseidon field}

\begin{figure}
\centering
\includegraphics[scale=0.3]{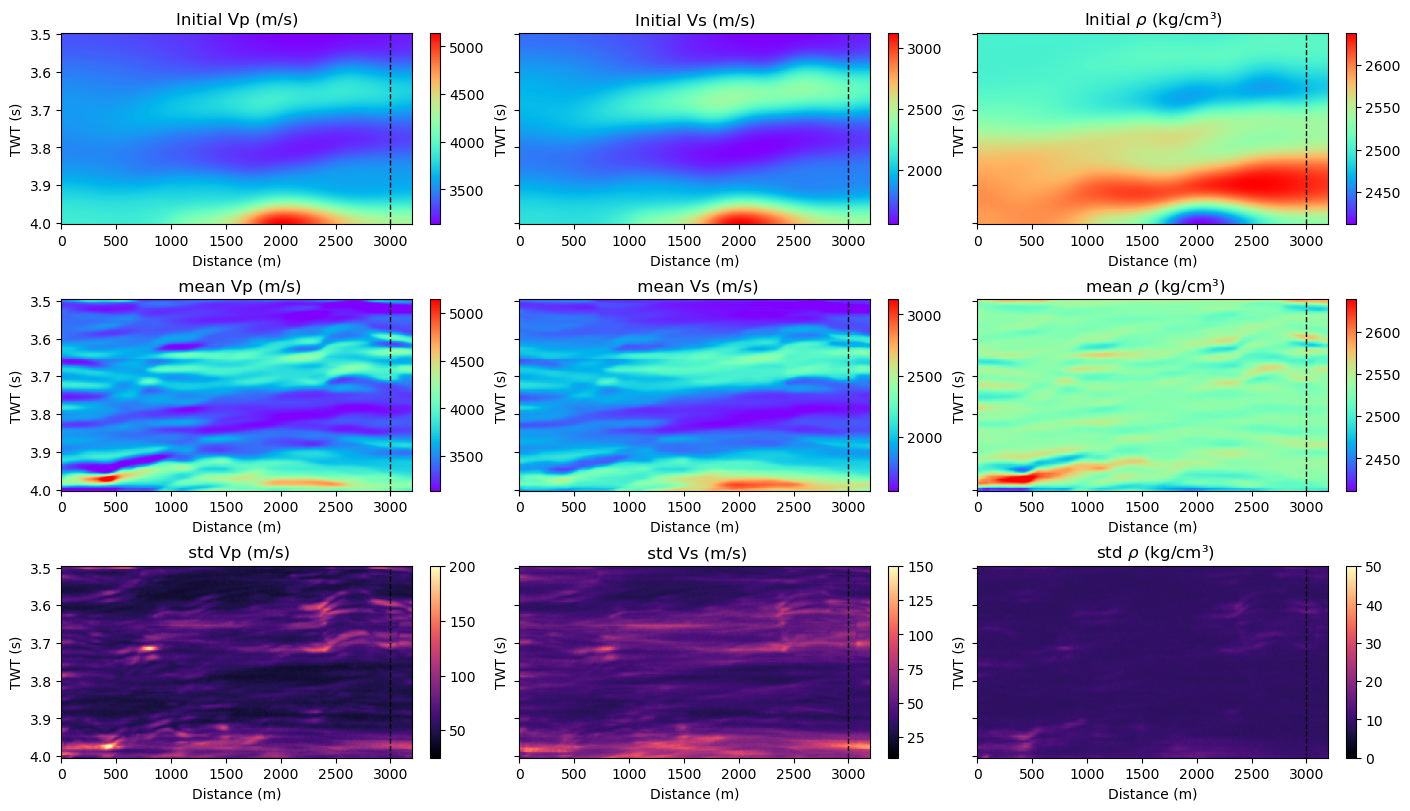} 
\caption{DPS-based posterior uncertainty quantification for the Poseidon field dataset, generated from an ensemble of 500 independent posterior realizations. (Top row) Initial smooth elastic models of $V_P$ (left), $V_S$ (middle), and $\rho$ (right) used to initialize the reverse diffusion process. (Middle row) Posterior mean estimates of $V_P$, $V_S$, and $\rho$ recovered by the DPS-guided inversion. (Bottom row) Posterior standard deviation of $V_P$, $V_S$, and $\rho$. The dashed vertical line marks the well location at $\approx$3000~m used as a reference 
for validation.} 
\label{fig11} 
\end{figure}

Having evaluated the DPS-based uncertainty quantification on the controlled Otway synthetic benchmark, we now apply the same framework to the Poseidon field dataset to assess its behaviour under realistic noise conditions and complex geology, with the analytical Bayesian solution of \citet{buland2003bayesian} serving as reference for comparison. Figure~\ref{fig11} presents the DPS-based uncertainty quantification results for the Poseidon field, generated from an ensemble of 500 independent posterior realizations over the deep gas-reservoir interval (TWT 3.5--4.0~s, distance 0--3000~m). The initial smooth models (top row) capture only the broad, low-frequency elastic trends for all three parameters, with $V_P$ ranging from $\approx$3500 to 5000~m/s, $V_S$ from $\approx$1500 to 3000~m/s, and $\rho$ 
from $\approx$2440 to 2600~kg/m\textsuperscript{3}, and a prominent low-velocity anomaly visible at $\approx$3.9--4.0~s TWT between 1500--2500~m lateral distance, consistent with the expected gas-saturated reservoir interval. The posterior mean (middle row) substantially improves upon the initial model, recovering fine-scale lateral layering and sharper lithological boundaries for $V_P$ and $V_S$, while the $\rho$ mean reveals a coherent high-density anomaly ($\approx$2580-2600~kg/m\textsuperscript{3}) at the base of the section near the well location, consistent with the underlying reservoir seal. The posterior standard deviation maps (bottom row) reveal a spatially structured uncertainty field concentrated along reflector interfaces and within the deep reservoir zone: $V_P$ uncertainty 
reaches up to $\approx$200~m/s at key layer boundaries, $V_S$ uncertainty peaks at 
$\approx$150~m/s over similar intervals, while $\rho$ uncertainty remains markedly low throughout the section (up to $\approx$50~kg/m\textsuperscript{3}), reflecting the inherent insensitivity of seismic reflection data to density in the field setting.

\begin{figure}
\centering
\includegraphics[scale=0.3]{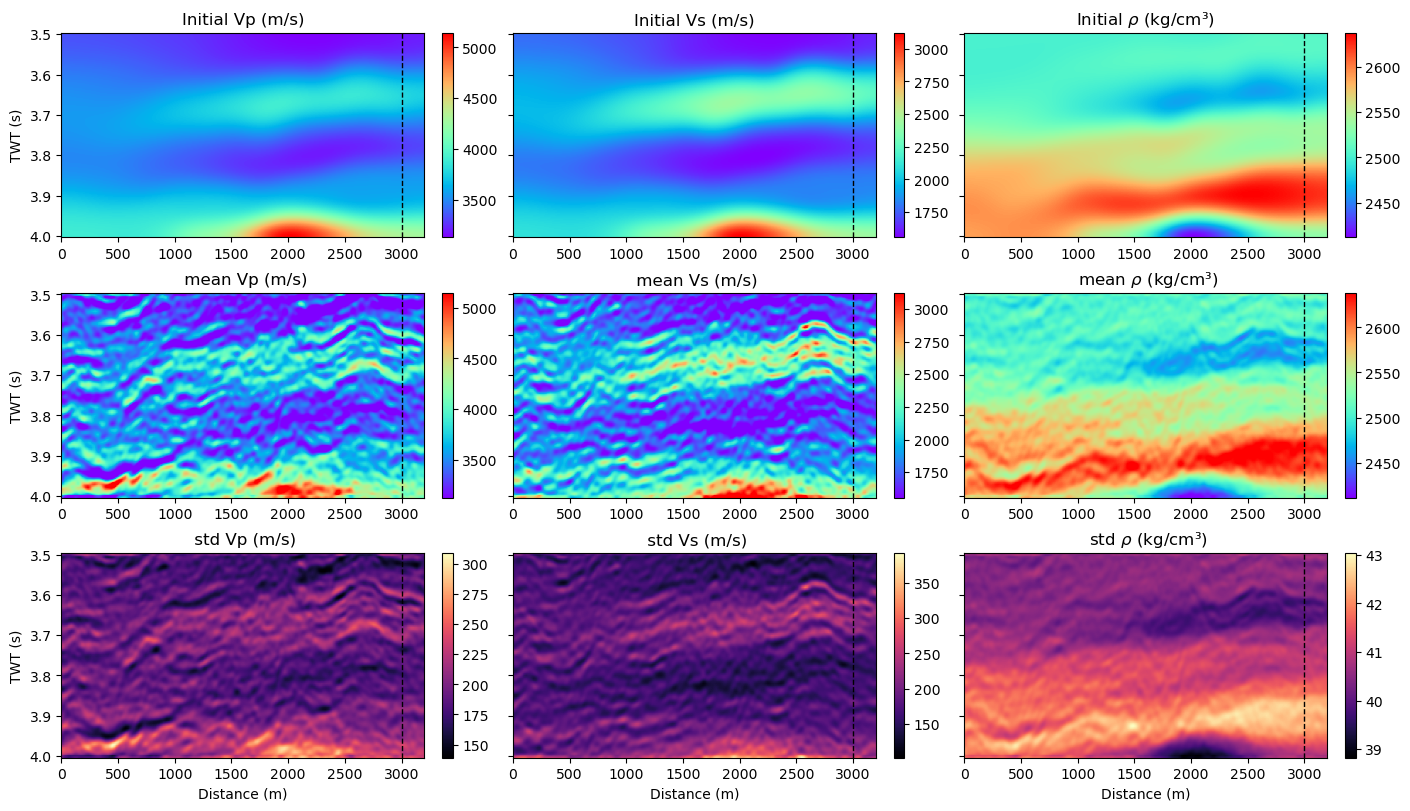} 
\caption{Analytical Bayesian posterior uncertainty quantification for the Poseidon field data. (Top row) Initial smooth elastic models of $V_P$ (left), $V_S$ (middle), and $\rho$ (right). (Middle row) Posterior mean estimates of $V_P$, $V_S$, and $\rho$ derived from the closed-form solution. (Bottom row) Posterior standard deviation of $V_P$, $V_S$, and $\rho$, representing the exact Gaussian posterior covariance. The dashed vertical line marks the well location at $\approx$3000~m.} 
\label{fig12} 
\end{figure}

Figure~\ref{fig12} presents the corresponding analytical Bayesian 
uncertainty results for the same Poseidon field interval. The posterior mean (middle row) recovers a comparable level of structural detail to the DPS mean, with fine-scale lateral  layering visible across all three parameters and a consistent high-density anomaly in the deep reservoir zone ($\approx$3.8--4.0~s TWT). The most noticeable difference, however, lies in the posterior standard deviation maps (bottom row): the analytical $V_P$ uncertainty reaches up to $\approx$300~m/s, the $V_S$ uncertainty peaks at $\approx$350~m/s, and the density uncertainty spans a narrow but elevated and spatially uniform range of $\approx$39--43~kg/m\textsuperscript{3} — all substantially broader than their DPS counterparts of $\approx$200~m/s, $\approx$150~m/s, and $\approx$50~kg/m\textsuperscript{3}, respectively. Furthermore, while the DPS standard deviation maps exhibit interface-concentrated uncertainty patterns, the analytical uncertainty maps are more spatially diffuse and laterally uniform, reflecting the global nature of the posterior covariance derived under the Gaussian prior assumption rather than the localized, data-misfit-driven collapse characteristic of the DPS ensemble. This consistent discrepancy between DPS and analytical uncertainty estimates on the Poseidon field data exhibiting the same nature observed for the Otway synthetic benchmark, reinforcing that the narrower DPS uncertainty bounds are a systematic property of the guidance-based sampling mechanism rather than an artifact of the synthetic test conditions.

\begin{figure}
\centering
\includegraphics[scale=0.30]{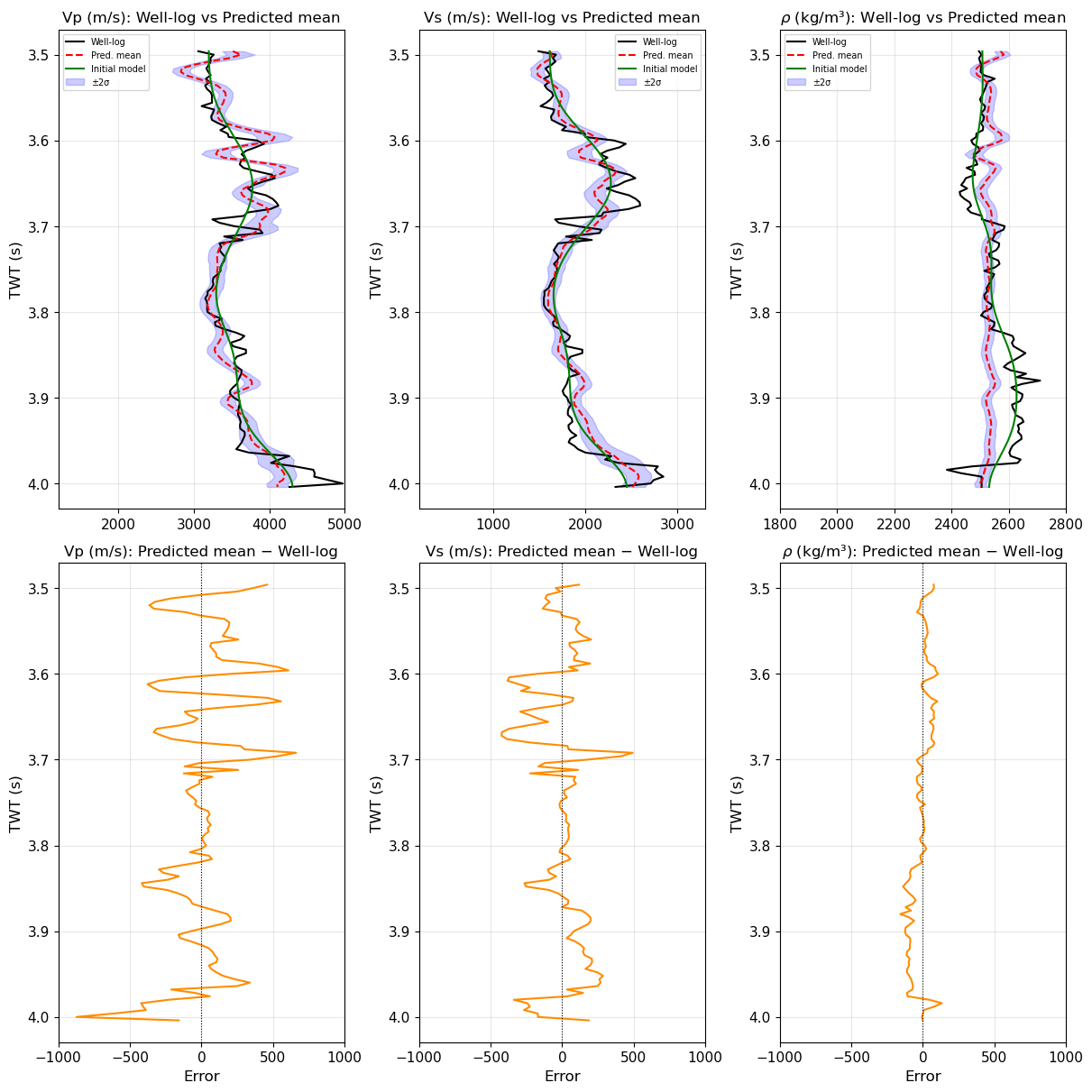} 
\caption{1D comparison of DPS-based posterior uncertainty at the well location for the Poseidon field dataset. (Top row) Vertical profiles of $V_P$ (left), $V_S$ (middle), and $\rho$ (right) as a function of TWT, showing the well-log (black solid), initial smooth model (green solid), posterior mean (red dashed), and the $\pm2\sigma$ uncertainty envelope (blue shading) derived from 500 independent DPS posterior realizations. (Bottom row) Pointwise error profiles (posterior mean $-$ well-log) for each elastic parameter, with the zero-error reference indicated by the dotted vertical line.} 
\label{fig13} 
\end{figure}

\begin{figure}
\centering
\includegraphics[scale=0.30]{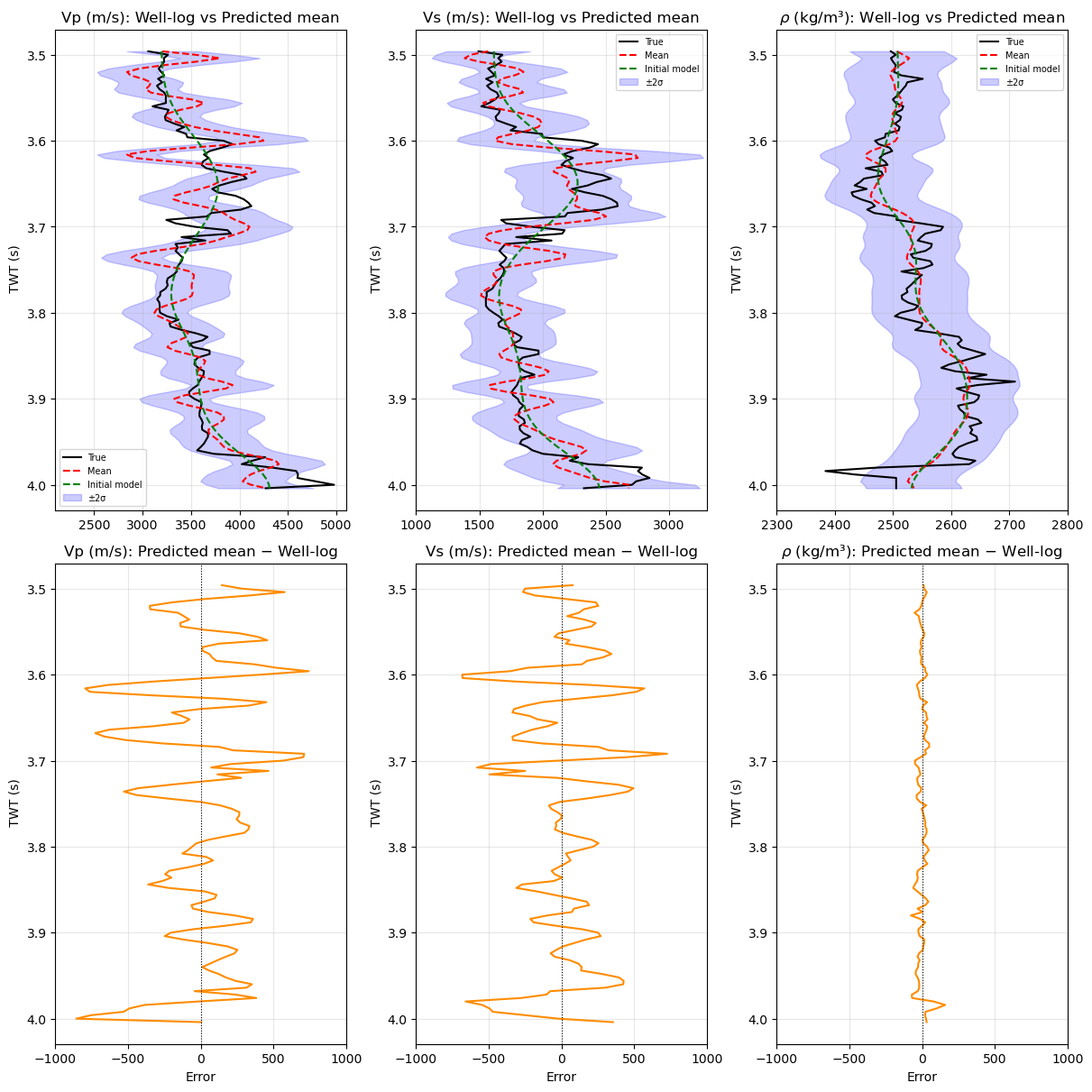} 
\caption{1D comparison of analytical Bayesian posterior uncertainty at the well location for the Poseidon field dataset. (Top row) Vertical profiles of $V_P$ (left), $V_S$ (middle), and $\rho$ (right) as a function of TWT, showing the true well-log (black solid), initial smooth model (green dashed), posterior mean (red dashed), and the exact $\pm2\sigma$ uncertainty envelope (blue shading) derived from the closed-form posterior covariance. (Bottom row) Pointwise error profiles (posterior mean $-$ well-log) for each elastic parameter, with the zero-error reference indicated by the dotted vertical line.} 
\label{fig14} 
\end{figure}

The 1D profiles extracted from the well location (Figures~\ref{fig13} 
and~\ref{fig14}) further corroborate the observations from the 2D 
uncertainty maps. For the DPS-based inversion (Figure~\ref{fig13}), the 
$\pm2\sigma$ envelope is markedly narrow, and the true well-log frequently falls outside the uncertainty bounds for $V_P$ and $V_S$, particularly in the deeper interval between 3.8--4.0~s TWT where structural complexity is highest, while the density envelope remains tightly constrained with near-zero pointwise errors throughout. For the analytical Bayesian method (Figure~\ref{fig14}), the $\pm2\sigma$ envelope is substantially wider and encompasses the true well-log for $V_P$ and $V_S$ across most of the TWT range, with the density envelope similarly containing the well-log almost everywhere. Both methods produce posterior means of comparable accuracy relative to the well-log, with similar error magnitudes across all three parameters, confirming that the key distinction lies not in the quality of the point estimate but in the reliability of the uncertainty quantification.

\section{Discussion}

The results presented in this work demonstrate that integrating a physics-based likelihood operator within a diffusion-based generative prior is effective for seismic elastic parameter estimation. Conventional deterministic approaches, while computationally efficient, are fundamentally limited by their ability to capture the complex joint distribution among $V_p$, $V_s$, and $\rho$ that characterizes realistic subsurface models. The diffusion prior trained on high-resolution elastic models drawn from diverse geological settings effectively encodes this joint distribution and guides the reverse diffusion process toward geologically consistent solutions, which is clearly reflected in the superior SSIM scores. Although DPS achieves superior structural reconstruction quality, its SNR is in some cases slightly lower than that of deterministic methods. This behavior is mainly due to the stronger sensitivity of SNR to local amplitude mismatches and high-frequency variations. In contrast, smoother deterministic reconstructions may artificially increase SNR by suppressing high-frequency details and reducing point-wise residual energy, despite producing less resolved and geologically less consistent models. The successful application to the Poseidon field dataset further underscores the practical viability of the approach under realistic conditions of noise, limited well control, and structural complexity. \\
It is important to emphasize that the overall DPS reconstruction quality is strongly influenced by the prior distribution learned by the diffusion model, which is itself determined by the training data distribution. If the training dataset does not adequately represent the geological characteristics of the target area, the resulting prior may introduce biases and potentially lead to less accurate reconstructions compared to conventional deterministic approaches. On the other hand, in practical industrial settings, companies often have access to additional geological and geophysical information, including regional studies, well logs, and multiphysics measurements. Such information can be incorporated into the construction of more representative training datasets and priors, enabling the diffusion model to constrain the inversion toward solutions that are both more accurate and more geologically consistent.\\
Despite the significant computational cost associated with the offline training stage of the diffusion model, the subsequent inference process can be efficiently performed on modern GPUs. In practice, the reconstruction time during inference becomes comparable to that of conventional deterministic inversion methods executed on CPUs, while providing improved structural fidelity and geological consistency.\\
The uncertainty analysis revealed an important behavior of the DPS framework: while the stochastic nature of the reverse diffusion process supports ensemble-based uncertainty quantification through multiple independent posterior realizations, the resulting uncertainty bands are systematically overconfident relative to the exact posterior covariance derived from the analytical Bayesian solution. This overconfidence stems from the data-misfit likelihood gradient embedded within DPS, which at each reverse diffusion step steers the evolving samples toward solutions that minimize the discrepancy between modeled and observed angle-stack seismic data, progressively collapsing the ensemble around low-misfit solutions and suppressing exploration of the posterior distribution. Combined with the learned prior, which further concentrates solutions toward high-probability regions of the training distribution, the resulting uncertainty bands become narrower than the true posterior spread.
In general, the estimated posterior standard deviation obtained with DPS is approximately one order of magnitude smaller than the reconstructed mean model, and in some cases up to two orders of magnitude smaller. This behavior is qualitatively consistent with the uncertainty estimates reported by \citet{ravasi2025}, where the posterior standard deviations are also generally observed to be approximately one order of magnitude smaller than the corresponding mean models. Nevertheless, in our case, the results still suggest that the uncertainty associated with DPS is underestimated. In particular, the strong variability observed in \citet{ravasi2025} when changing the training datasets and diffusion sampling strategies indicates that the uncertainty estimate results are once again highly sensitive to the learned prior distribution encoded within the diffusion model. We therefore believe that a substantial portion of the observed uncertainty behavior is controlled by the prior itself, and a more systematic investigation of this dependency will be an important direction for future work.\\
Despite these limitations, the proposed framework demonstrates the practical feasibility of obtaining fast two-dimensional uncertainty quantification directly in the model domain, rather than through independent trace-wise analytical approximations. Furthermore, DPS naturally generates high-resolution posterior realizations that preserve realistic geological structures, a characteristic that is difficult to achieve with many conventional uncertainty quantification approaches \citet{ravasi2025}. Addressing the current discrepancy through improved likelihood approximations, annealed guidance schedules, or hybrid analytical-diffusion posterior formulations remains an important direction for future research.

\section{Conclusions}

In this work, we presented a unified diffusion-based inversion framework for the simultaneous recovery of $V_P$, $V_S$, and $\rho$ from angle-stack seismic data using DPS. By embedding a linearized forward operator based on the Aki-Richards approximation as the likelihood guidance term within the reverse diffusion process, the framework achieves physically-consistent reconstruction without task-specific retraining of the diffusion prior. The unsupervised prior, learned from a dataset of high-resolution elastic models, captures the complex non-Gaussian joint distribution of the three elastic parameters, providing geologically realistic reconstructions. Validation on the 2D Otway synthetic benchmark demonstrated superior reconstruction accuracy over both LSQR and ADMM-based total variation inversion baselines, particularly in resolving sharp lithological contrasts and laterally heterogeneous structures. Application to field data from the Poseidon field, NW-Shelf, Browse Basin, Australia, confirmed the practical viability of the framework under realistic noise conditions and limited well control. Uncertainty quantification through multiple independent posterior realizations, benchmarked against the closed-form analytical Bayesian solution, revealed that DPS uncertainty estimates are systematically overconfident, as the data-misfit likelihood gradient and the concentrated diffusion prior together suppress posterior spread relative to the exact posterior covariance. Although,DPS outperforms LSQR and ADMM+TV in recovering better structural features, conventional methods marginally surpass DPS in peak signal-to-noise ratio for $V_P$ and $\rho$ in certain intervals (in the synthetic example case), suggesting that the diffusion prior can occasionally over-regularize fine-scale amplitude variations toward the training distribution at the expense of data-consistent high-frequency detail. The framework also inherits the limitations of the linearized Aki-Richards approximation, restricting its reliability to moderate-contrast interfaces and moderate incidence angles. Addressing these shortcomings through improved likelihood approximations, annealed guidance schedules, and replacing the linearized Aki-Richards operator with the full Zoeppritz equations as the guidance term would extend the framework to wide-angle reflections and strong-contrast interfaces, further broadening its applicability to challenging exploration targets.

\section*{Acknowledgments}
This publication is based on work supported by the King Abdullah University of Science and Technology (KAUST). The authors are grateful for support from DeepWave sponsors and the computational resources provided by KAUST’s high-performance
computing facilities.




\end{document}